\documentclass[sigconf]{acmart}
\usepackage{times}  
\usepackage{helvet}  
\usepackage{courier}  
\usepackage{url}  
\usepackage{graphicx} 
\usepackage{natbib}  
\usepackage{caption} 
\usepackage{algorithm}
\usepackage{algorithmic}

\usepackage{url}
\usepackage[utf8]{inputenc}
\usepackage[switch]{lineno}
\usepackage{subfig}
\usepackage{braket}
\usepackage{threeparttable}
\usepackage{array}
\usepackage{color}
\usepackage{comment}
\usepackage{amsmath}
\usepackage{mathtools}
\usepackage{amsthm}
\usepackage{newfloat}
\usepackage{listings}
\DeclareCaptionStyle{ruled}{labelfont=normalfont,labelsep=colon,strut=off} 
\floatstyle{ruled}
\newfloat{listing}{tb}{lst}{}
\floatname{listing}{Listing}
\AtBeginDocument{%
  }

\renewcommand\footnotetextcopyrightpermission[1]{}
\setcopyright{acmlicensed}
\copyrightyear{2025}
\acmYear{2025}
\acmDOI{XXXXXXX.XXXXXXX}
\acmConference[QNLP '25]{Quantum AI and NLP 2025 Conference}{August 06--08,
  2025}{Indiana University, Bloomington, Indiana, USA}

\begin{document}

\title[SA-DQAS]{SA-DQAS: Integrating Differentiable Quantum Architecture Search with Transformers for Enhanced Variational Quantum Algorithms}

\author{Yize Sun}
\email{yize.sun@campus.lmu.de}
\orcid{0009-0007-2921-2858}
\affiliation{%
  \institution{LMU \& Siemens AG \& MCML}
  \city{Munich}
  \state{Bavaria}
  \country{Germany}
}

\author{Jiarui Liu}
\email{jiarui.liu@campus.lmu.de}
\affiliation{%
  \institution{LMU}
  \city{Munich}
  \state{Bavaria}
  \country{Germany}
}

\author{Zixin Wu}
\email{wzx19990414@gmail.com}
\affiliation{%
  \institution{LMU}
  \city{Munich}
  \state{Bavaria}
  \country{Germany}
}

\author{Volker Tresp}
\email{volker.tresp@lmu.de}
\affiliation{%
  \institution{LMU}
  \city{Munich}
  \state{Bavaria}
  \country{Germany}
}

\author{Yunpu Ma}
\email{cognitive.yunpu@gmail.com}
\affiliation{%
  \institution{LMU}
  \city{Munich}
  \state{Bavaria}
  \country{Germany}
}








\renewcommand{\shortauthors}{Sun et al.}


\begin{abstract}
We introduce SA-DQAS, a novel framework that enhances Differentiable Quantum Architecture Search (DQAS) by integrating a self-attention mechanism, enabling more effective quantum circuit design for variational quantum algorithms. Unlike DQAS, which treats placeholders independently, SA-DQAS captures inter-placeholder dependencies to improve architecture learning. We evaluate SA-DQAS across multiple tasks, including MaxCut, Job-Shop Scheduling Problem (JSSP), quantum chemistry simulation, and error mitigation. Experimental results show that SA-DQAS outperforms baselines and prior QAS methods in most cases, producing architectures with better stability, convergence, and noise resilience. To assess scalability and hardware readiness, we further test SA-DQAS-generated circuits on IBM's quantum device using the MaxCut problem. Circuits trained on small graphs are stacked to solve larger instances without retraining, demonstrating generalization to real hardware and larger problem sizes. Our results suggest that SA-DQAS not only improves circuit quality during training but also enables practical deployment on near-term quantum devices. This research represents the first successful integration of self-attention mechanism with DQAS.
\end{abstract}




\maketitle

\section{Introduction}
In recent years, quantum computing (QC) has advanced rapidly, achieving notable progress in areas such as image classification, circuit architecture search, quantum reinforcement learning, knowledge graph embedding, and approximate optimization problems~\cite{alam2021quantum, zhang2022differentiable, wu2023quantumdarts, ma2019variational, amaro2022case, kim2025quantum, chen2024quantumqtrain}. However, in the noisy intermediate-scale quantum (NISQ) era, quantum algorithms face hardware limitations that cause performance discrepancies on noisy devices. Variational quantum algorithms (VQAs) have emerged as a leading strategy in this era~\cite{cerezo2021variational}, utilizing parameterized quantum circuits (PQCs) with trainable gates optimized by classical algorithms. The success of VQAs, however, heavily relies on circuit architecture.

Quantum architecture search (QAS) automates the design of task-specific quantum circuits, aiming to create efficient and high-performing architectures. Various QAS algorithms have been developed, including gradient-based~\cite{zhang2022differentiable, wu2023quantumdarts}, reinforcement learning-based~\cite{dai2024quantum, giovagnoli2023qneat}, predictor-based~\cite{he2023gsqas}, and predictor-free~\cite{sun2024quantum} methods. Differentiable quantum architecture search (DQAS) is a gradient-based framework that optimizes a super-circuit by learning architecture parameters. While effective, DQAS treats each circuit placeholder independently, ignoring potential dependencies across the circuit.

In this work, we propose \textbf{SA-DQAS}, which integrates a self-attention mechanism into DQAS. This extension allows the architecture search process to capture inter-placeholder relationships, thereby improving stability and generalization.

We evaluate SA-DQAS across multiple tasks, including MaxCut, Job-Shop Scheduling Problem (JSSP), error mitigation, and quantum chemistry simulation. Extensive simulations show that SA-DQAS outperforms baselines and prior QAS methods in most cases, though not universally. Furthermore, we explore the generalization and scalability of architectures found by SA-DQAS by conducting additional experiments on real quantum hardware. In these experiments, we first search for a compact circuit on an 8-qubit MaxCut instance and then stack the learned circuit to solve 16-qubit problems. We test the resulting architectures on IBM’s \texttt{ibmq\_sherbrooke} device, showing that the trained blocks generalize to larger graphs and noisy devices without retraining.

Our contributions are summarized as follows:
\begin{enumerate}
    \item We introduce SA-DQAS, a novel QAS algorithm that incorporates a self-attention mechanism to enhance architecture parameter learning.
    \item We demonstrate that SA-DQAS achieves competitive or superior performance across a variety of tasks—including MaxCut, JSSP, quantum chemistry, and fidelity tests under noise—through comparisons with multiple baseline methods.
    \item We investigate the scalability and transferability of learned circuits by stacking blocks trained on small graphs to solve larger instances, validated on a real quantum device.
\end{enumerate}

\section{Related Work}
\label{Background and Related Work}

\subsection{DQAS}
Various quantum architecture search algorithms can automatically search for near-optimal quantum circuit architectures~\cite{du2022quantum}. Some approaches define a parameterized super-circuit with multiple layers and iteratively train this super-circuit by sampling and updating parameters~\cite{du2022quantum,meng2021quantum,zhang2022differentiable,chen2024differentiable,chen2025differentiable}. Genetic algorithms~\cite{ding2022evolutionary,ding2023multi,chen2024evolutionary} evolve circuits over time, while RL-based algorithms~\cite{fosel2021quantum,kuo2021quantum,ye2021quantum} train an agent to select circuit components. DQAS~\cite{zhang2022differentiable} simultaneously trains weights and the circuit architecture, whereas neural predictor-based approaches~\cite{zhang2021neural} build machine learning models based solely on network structure.

DQAS constructs a circuit architecture by placing operations in each placeholder, utilizing a predefined operation pool $\mathcal{O}$, a circuit with $p$ placeholders, an architecture parameter $\alpha$, and weights $\theta$. The distribution of each circuit architecture candidate is approximated using a probabilistic model, where the probability of each candidate is the normalized architecture parameter $\alpha$.

Rather than evaluating all possible architecture candidates, DQAS samples a batch of candidates and calculates their local losses based on task-specific objective functions. It then updates $\alpha$ and $\theta$ simultaneously using a gradient-based optimizer, based on the global loss $\mathcal{L}$ and local loss $L$, as given by:
\begin{equation}\label{Eq.Loss}
    \mathcal{L}=\sum_{k\in P(k,\alpha)}\frac{P(k,\alpha)}{\sum_{k'\in P(k,\alpha)}
    P(k',\alpha)}L(U_k) \quad ,
\end{equation}
where
\begin{align}\label{Eq.prob}
    P(k,\alpha) &= \prod_{i=1}^{p}p(k_i,\alpha_i), \\
    L(U_k) &= \sum_i \left(\bra{\psi_i}U_k^{\dagger}OBS\, U_k\ket{\psi_i} - y_i\right)^2
\end{align}

The symbols $\psi_i$, $OBS$, and $y_i$ denote the input data, observable, and corresponding label, respectively, while $k$ determines the circuit structure from the probabilistic model. The local loss is computed by summing the individual losses over all input data points. In some cases—such as the fidelity or MaxCut problems—we consider only a single input ($i=1$) as a representative example.

\subsection{Job Shop Scheduling Problem}
JSSP  is a combinatorial optimization problem crucial in industrial production environments, where it aims to enhance efficiency. Due to its computational complexity, JSSP quickly becomes intractable for large problem sizes. Various quantum computing approaches, such as quantum gate models and quantum annealing, have been explored to solve JSSP. JSSP can be formulated as a Quadratic Unconstrained Binary Optimization (QUBO) problem, and Variational Quantum Algorithms (VQAs) are employed in quantum gate models to find optimal solutions. It has been demonstrated that algorithms like VQE, QAOA, VarQITE, F-VQE, particularly F-VQE, can effectively solve JSSP instances with up to 23 qubits~\cite{amaro2022case}. Quantum annealing is emerging as a promising approach for solving combinatorial optimization problems like JSSP~\cite{carugno2022evaluating,venturelli2016job}, as it can handle larger-scale problems without being constrained by the number of qubits, unlike quantum gate models.

\subsection{Attention Mechanism}
Self-attention, as introduced in Transformer architectures~\cite{vaswani2017attention}, has become a key mechanism in modeling dependencies across sequence elements. It enables each position to attend to all others, leading to breakthroughs in fields like natural language processing and image generation. Recent studies have explored the application of attention to quantum models~\cite{chen2025quantum}, but not specifically in the context of architecture search.

The attention mechanism allows models to focus on the most significant parts of the input.
Absolute positional information can be learned as parameters during training. Relative positional embedding is another approach, as demonstrated in \cite{shaw2018self}. CNN-based models use a multi-step attention mechanism for each decoder layer \cite{gehring2017convolutional}, while RNN-based translation models apply the attention mechanism in the decoder and use bidirectional RNNs in the encoder \cite{bahdanau2014neural}. The BERT model operates in two stages: pre-training and fine-tuning \cite{devlin2018bert}. 
\section{Methodology}
\begin{figure*}
    \centering
    \includegraphics[width=\linewidth]{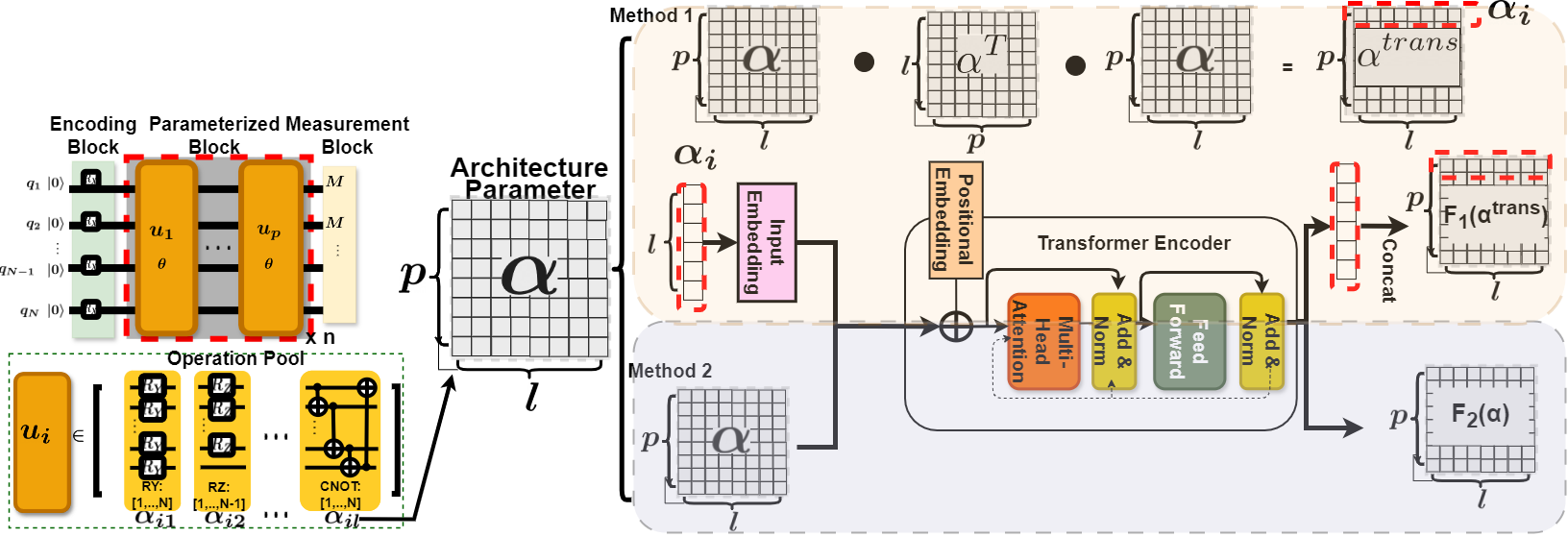}
    \caption{The illustration of the proposed framework SA-DQAS.}
    \Description{
        On the left side is the common part of both algorithms (Step $1$ and Step $2$). On the right side, we show the differences between both algorithms. SA-DQAS uses the attention modules $F_1$ or $F_2$ to improve the architecture parameters $\alpha$, such that each operation shares positional and fitness information among operation candidates in different placeholders.}
    \label{Fig.SA-DQAS}
\end{figure*}

Our work SA-DQAS derives from integrating self-attention mechanism with DQAS, which uses a gradient-based method to construct a quantum circuit by replacing these placeholders with one possible candidate operation from the operation pool. We show our algorithm in Fig~\ref{Fig.SA-DQAS}. 
We denote the operation pool as $\mathcal{O}$. A circuit is a sequence of $p$ placeholders described as
\begin{flalign}
    U=\prod_{i=0}^p u_i(\theta_i),
\end{flalign} where $u_i$ stand for unitary placeholder, and $\theta_i$ are the parameters of trainable gates. The architecture parameter matrix is $(\alpha_{ij}) \in \mathbb{R}^{p\times l}$ with $l=|\mathcal{O}|$. The probability is calculated by Eq~\ref{Eq.prob} 
. This setting makes the search process continue and focuses on operations in one placeholder but ignores the relationship among operation candidates placed on different placeholders.


\paragraph{Encoder $F1$\&$F2$}
As shown in Fig~\ref{Fig.SA-DQAS}, each architecture parameters vector $\alpha_{i}$ is viewed as a word, and the architecture matrix $\alpha$ is viewed as a sentence. The output of the encoder is the architecture parameters, including relationship features, and is added to the original architecture parameter matrix $\alpha$:
\begin{flalign}
    \alpha'=\alpha+\beta F(\alpha),
\end{flalign}
where $F$ is the encoder and $\beta$ is the scaling coefficient.

From different perspectives, by viewing a placeholder as a token, we propose one encoder with two types of input: $F_1$ and $F_2$.
Regarding the encoder $F_1$, we use sinusoidal position embedding $\alpha_i^{pos}$ for $N$ layers-encoder. Each layer consists of two sublayers:
\begin{flalign}
     &\alpha_i^{n_1} =\text{LayerNorm}(\alpha_i^a),\\
     &\alpha_i^{\text{a}} = \text{MultiHead}(\alpha_i^{pos},\alpha_i^{pos},\alpha_i^{pos}) + \alpha_i^{pos}.
\end{flalign}
The second sub-layer is a position-wise fully connected feed-forward network:
\begin{flalign}
    &\alpha_i^{n_2} = \text{LayerNorm}(\alpha^{a_2}_i),\\
    &\alpha^{a_2}_i = \alpha^{n_1}_i + g(\alpha^{n_1}_i),\\
    &g(\alpha^{n_1}_i) = W_2\text{max}(0, W_1\alpha^{n_1}_i+b_1)+ b_2.
\end{flalign}
And $\alpha$ is updated by:
\begin{align}\label{Eg.F1}
    &\alpha_{1}' = \alpha +\beta F_1(\alpha^{trans}),\\
    &\alpha^{trans} = \alpha \times \alpha^T \times \alpha.
\end{align}

For $F2$, we view the original architecture matrix $\alpha$ as an input for the encoder. The output of this method is 
\begin{flalign}
    \alpha_{2}' = \alpha + \beta F_2( \alpha ).
\end{flalign}
We expect that the self-attention mechanism can give us additional features without strongly reducing the effect of the original architecture parameters. As a result, we select a sufficiently small $\beta$ as a scaling coefficient.

\paragraph{Loss function}
The loss function for the whole process of circuit architecture search is described in~Eq.\ref{Eq.Loss}. In order to get a gradually fixed architecture parameter, we select the distance of the output of the encoder between the current training step and one previous training step as our loss metric. For each training step $t$, the encoder parameters are updated by 
\begin{flalign}
    &w_{t} = w_{t}-\eta \Delta_w L_{\text{encoder}}\\
    &L_{\text{encoder}} = \max_{1\leq i \leq p}\max_{1\leq j \leq l}|F_{t-m}-F_t(\alpha^{trans}_{ij})|,
\end{flalign}
where $F_t$ is the output matrix of $F$ in training step $t$. We set $m=1$ in our experiment and $F(\alpha^0)$ is initialized by $0$.

\section{Experiments}
\subsection{Experiment settings}
In our experiments, we evaluate our framework using the JSSP, the Max-cut problem, quantum chemistry and fidelity measurements. We define four different types of operation pools, $\mathcal{O}_1\mbox{-}\mathcal{O}_4$, which are used in the Max-cut problem, JSSP and quantum chemistry problem. The basic training settings and the specific types of gates included in $\mathcal{O}_1\mbox{-}\mathcal{O}_4$ are given in Table~\ref{Tab:op setting}. 

\newcolumntype{C}[1]{>{\centering\arraybackslash}p{#1}}
Operation pools:
 \begin{table}[htbp]
     \centering
     \caption{The types of gates in operation pools $\mathcal{O}_1\mbox{-}\mathcal{O}_4$}
     \begin{threeparttable}
         \begin{tabular}[width=1.0\linewidth]
         {cc}
             \hline Operation pool & Gate type \\
             \hline $\mathcal{O}_1$  & \texttt{RY}, \texttt{RZ}, \texttt{H}, \texttt{CZ}, \texttt{I} \\
                    $\mathcal{O}_2$  & \texttt{RY}, \texttt{RZ}, \texttt{H}, \texttt{CNOT}, \texttt{I}  \\
                    $\mathcal{O}_3$ & \texttt{RY}, \texttt{RZ}, \texttt{H}, \texttt{CZ}, \texttt{CNOT}, \texttt{I}  \\
                    $\mathcal{O}_4$ & \texttt{H}, \texttt{U3}, \texttt{CU3}, \texttt{I}  \\
             \hline      
         \end{tabular}
     \end{threeparttable}
     \label{Tab:op setting}
 \end{table}
The smallest and largest operation pools of $\mathcal{O}_4$ with five qubits in the JSSP for each type are shown as follows:
\begin{equation}
    \begin{aligned}
        {\rm op4\mbox{-}1} = \{&\texttt{U3}:[0,1,2,3,4], \\
        &\texttt{U3}:[0,1,2,3], \texttt{U3}:[1,2,3,4], \\
        &\texttt{U3}:[0,1,2], \texttt{U3}:[1,2,3], \texttt{U3}:[2,3,4], \\
        &\texttt{U3}:[0,1], \texttt{U3}:[1,2], \texttt{U3}:[2,3], \texttt{U3}:[3,4], \\
        &\texttt{CU3}:[0,1,2,3], \texttt{E}:[0,1,2,3,4]\} \\
        {\rm op4\mbox{-}4} = \{&\texttt{U3}:[0,1,2,3,4], \texttt{H}:[0,1,2,3,4],\\
        &\texttt{CU3}:[0,1,2,3],\texttt{E}:[0,1,2,3,4]\}
    \end{aligned}
\end{equation}
The last operation \texttt{E} in the operation pools refers to an identical gate. We adjust the size of the operation pools according to the following rule: each time we reduce the size, we remove all single-gate operations with the current smallest working range. For example, the largest operation pool of $\mathcal{O}_4$ includes the \texttt{U3} gate, which has a working range length of 2. The smallest pool contains single-qubit gates with a working range length of 5. Consequently, for each type, with $N$ qubits, we create $N-1$ different sizes of operation pools, numbered in order from largest to smallest. 

In addition to the above tasks, we also include a new experiment on real quantum hardware to assess the scalability and practical robustness of our searched circuit architectures. For this evaluation, we focus on the MaxCut problem due to its flexibility in graph size and its widespread use as a quantum benchmark.

In this hardware experiment, we adopt a new operation pool: $\mathcal{O}_{\text{scale}} = \{\texttt{CNOT}, \texttt{RY}, \texttt{RZ}\},$
which provides a minimal yet expressive set of operations suitable for NISQ hardware execution.

To test architectural scalability, we first use SA-DQAS to search for a performant circuit architecture on a small 8-node MaxCut graph. The learned architecture is then transferred to larger problem instances by stacking the learned circuit blocks. Specifically, we evaluate performance on 16-node graphs by repeating the trained block multiple times. This setup allows us to examine whether architectures discovered on small circuits can generalize to larger problems and maintain performance when deployed on real quantum devices.

\subsection{Max-cut problem}
\begin{figure}[!htbp]
    \centering
    \includegraphics[width=0.7\linewidth]{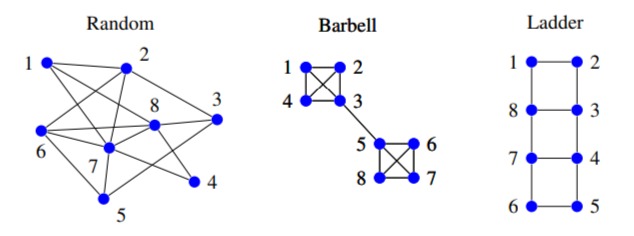}
    \caption{Benchmark graphs used for Max-cut problem.}
    \label{Fig.benckmark}
\end{figure}

In this section, we address the Max-cut problem using DQAS under various settings. 
The benchmark graphs, consisting of 8 nodes, along with the baseline circuit, are presented in Fig~\ref{Fig.benckmark}.

We previously discussed incorporating a self-attention mechanism into DQAS comparing with baseline and quantumDarts~\cite{wu2023quantumdarts}. SA-DQAS finds circuits using different operation pools, then use these circuits to solve the Max-cut problem on the three graphs. The operation pools op1-* to op4-* are derived from $\mathcal{O}_1$ to $\mathcal{O}_4$. Both controlled gates and single-qubit gates have a maximum working range of [0,1,2,3,4,5,6,7], leading to 28 distinct operation pools. We compare the effectiveness of selected circuits containing controlled gates in solving the Max-cut problem against the baseline. 

\begin{figure*}[!ht]
    \centering
    \subfloat[Max-cut Random]{
    \includegraphics[width=0.32\linewidth]{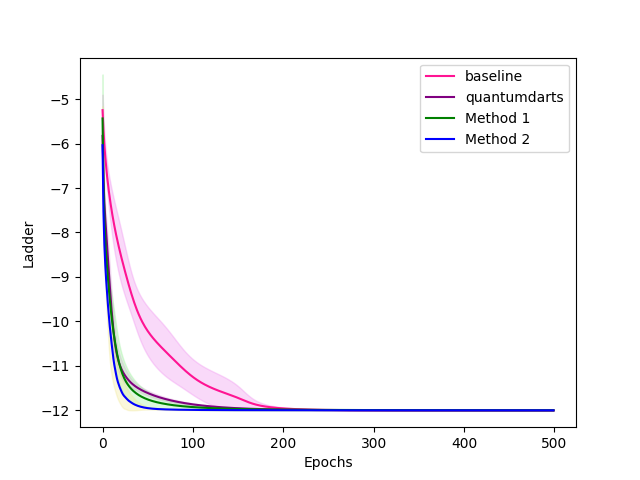}
    \label{fig: op3-6R}
    }   
    \subfloat[Max-cut Barbell]{
    \includegraphics[width=0.32\linewidth]{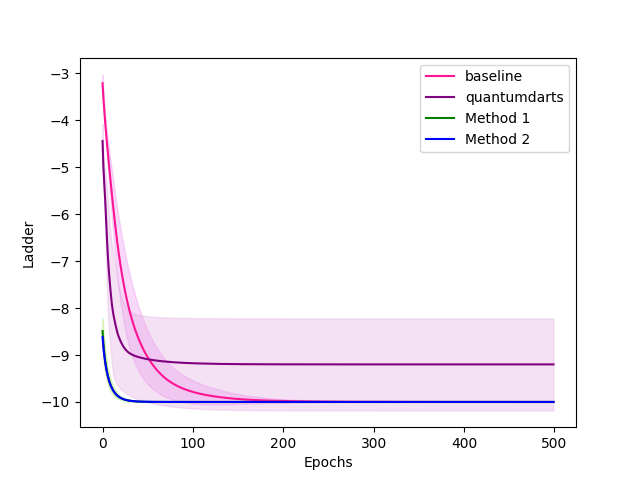}
    \label{fig: op2-7R}
    }
    \subfloat[Max-cut Ladder]{
    \includegraphics[width=0.32\linewidth]{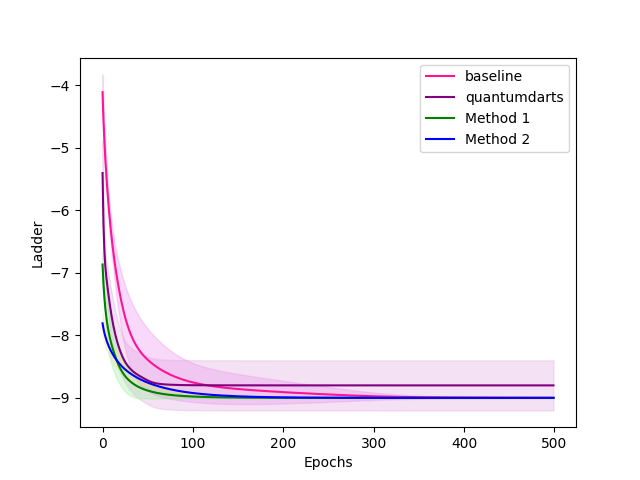}
    \label{fig: op3-4R}
    }
    \caption{Evaluation of the performance of circuits generated by SA-DQAS and quantumDarts. Method 1 uses the encoder $F_1$, and Method 2 uses the encoder $F_2$. The line represents the average results of 5 trials with different initial parameters. The SA-DQAS shows an much more stable performance for Max-cut problem comparing with quantumDarts.}
  \label{fig:BarbellRandom}
\end{figure*}

From the results in Fig. \ref{fig:BarbellRandom}, we observe that some circuits quickly converge to the minimum energy, outperforming the baseline, while others show slower convergence. Additionally, the shaded areas representing standard deviation for some circuits are much larger, and their average results fail to reach the exact minimum energy, indicating instability in their performance. In some trials, these circuits even reach incorrect solutions, likely due to the limitation of quantumDarts explained in their paper for Max-cut problem. However, all circuits from SA-DQAS converge to the minimum energy faster than the baseline, with no significant standard deviations, and all average results achieve the exact minimum energy.
\begin{figure*}[!ht]
    \centering
    \subfloat[W$<$WO op2-4$_4$]{
    \includegraphics[width=0.15\linewidth]{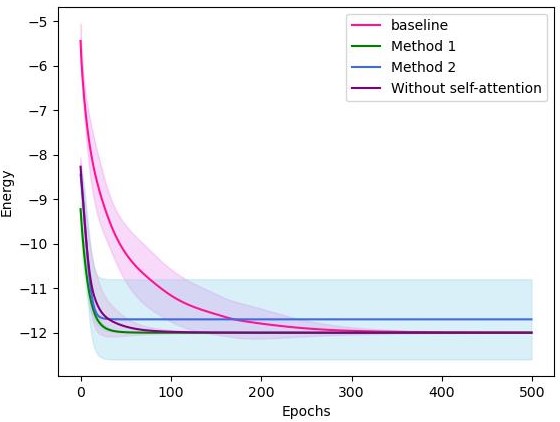}
    \label{fig: op2-4A1A2MH1}
    }   
    \subfloat[W$>$WO op4-7$_4$]{
    \includegraphics[width=0.15\linewidth]{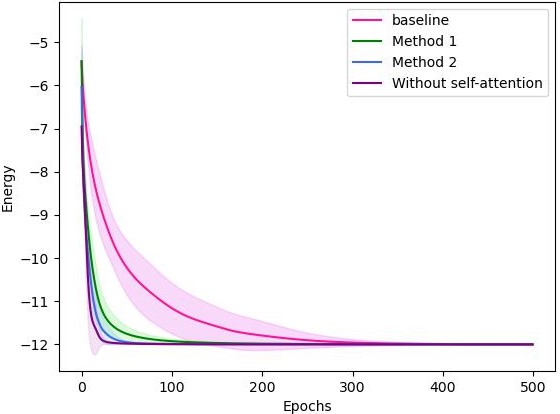}
    \label{fig: op4-7A1A2MH1}
    }
    \subfloat[W$\approx$WO op4-5$_4$]{
    \includegraphics[width=0.15\linewidth]{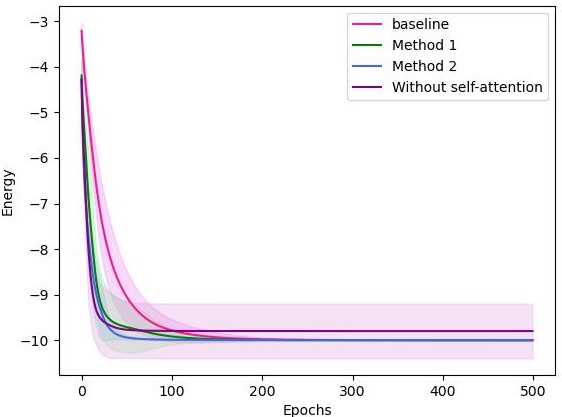}
    \label{fig: op4-5A1A2MH2}
    }   
    \subfloat[W$\approx$WO op3-3$_4$]{
    \includegraphics[width=0.15\linewidth]{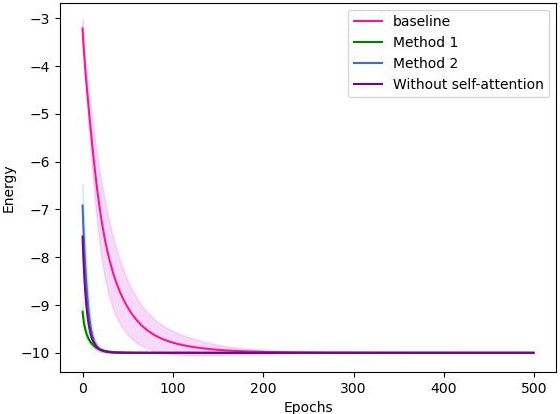}
    \label{fig: op3-3A1A2MH2}
    }    
    \subfloat[W$<$WO op3-1$_4$]{
    \includegraphics[width=0.15\linewidth]{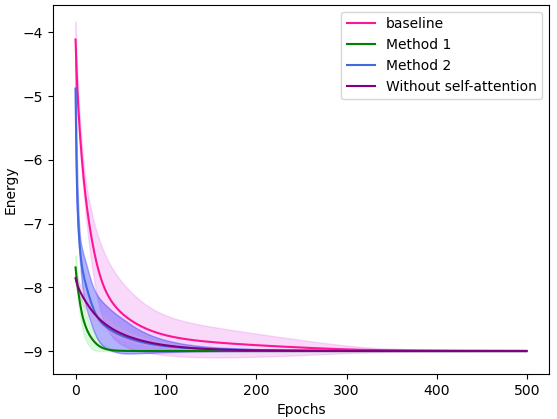}
    \label{fig: op3-1A1A2MH3}
    }   
    \subfloat[W$>$WO op3-6$_4$]{
    \includegraphics[width=0.15\linewidth]{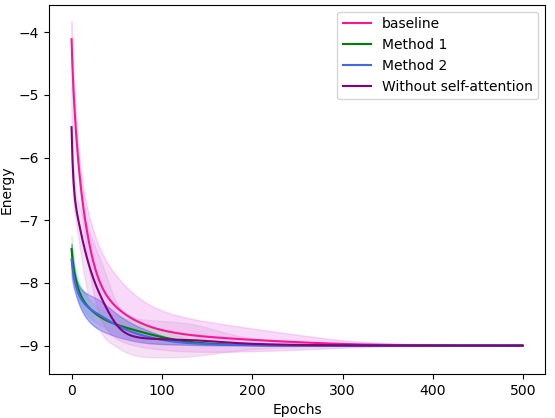}
    \label{fig: op3-6A1A2MH3}
    }
  \caption{Performance of circuits generated with (W) and without (WO) self-attention across different Max-cut problems. (a) and (b) show the evaluation results for Random graphs, (c) and (d) for Ladder graphs, and (e) and (f) for Barbell graphs. Circuits created using method 1 consistently converge faster to the minimum energy than those generated by method 2, as shown in (b), (c), and (e). While some circuits from both methods perform similarly, as seen in (d) and (f), method 2 circuits occasionally exhibit significant deviation, as in (a). Overall, circuits from both methods outperform the baseline in finding optimal solutions, with method 1 generally producing more stable and reliable results, less sensitive to initial parameter variations.}
  \label{fig:compareA1A2M}
\end{figure*}

\subsection{Experiments on JSSP}
In this section, we conduct experiments using SA-DQAS for the simplest JSSP task, as defined in \cite{amaro2022case}, with five qubits. The manually designed circuit shown in Fig. \ref{fig: baseline} and the experiment setting are from the paper~\cite{amaro2022case} and used as our baseline. In these experiments, all qubits are initialized to the state $\ket{0}$, and \texttt{rx} gates with a rotation angle of $\pi$ are applied to form the encoding block. There is only one parameterized block containing four placeholders, and all placeholders are updated at each learning step. The energies $E$ are scaled to the range $[0,1]$:
\begin{equation}
e=\frac{E-E_{\rm min}}{E_{\rm max}-E_{\rm min}} \in [0,1] \quad,
\end{equation}
where $E_{\rm min}$ and $E_{\rm max}$ are the minimum and maximum energies, respectively. When $e \approx 0$, the minimal energy is reached, indicating the optimal solution.
\begin{figure}[!ht]
    \centering
    \includegraphics[width=0.9\linewidth]{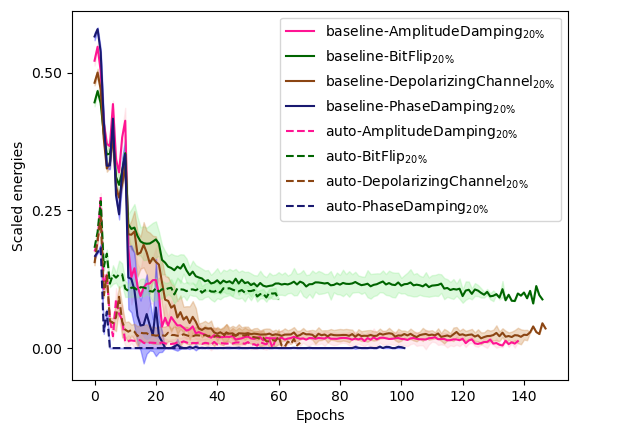}
    \caption{Performance evaluation in noisy environments: We assess the noise resistance of auto-generated circuits by introducing noise models that add 20\% of various noise types to each qubit in the top-performing generated circuit.}
    \label{fig:noise evaluation}
\end{figure}

In Fig. \ref{fig:noise evaluation}, the shaded standard deviation areas of the designed architectures are significantly smaller than those of the baseline, indicating that the learned circuits produce more consistent outcomes. Under BitFlip noise, both the baseline and the searched circuit yield average results that are considerably higher than the minimum energy. However, the ASP and minimum energies achieved by the designed circuits outperform the baseline across different models. Although neither the learned circuits nor the baseline consistently find the exact minimum energy in some trials due to the influence of noise, the automatically designed circuits are generally more noise-resilient and reliable.
\begin{figure}[!ht]
    \centering
    \subfloat[Comparison of placeholders\label{fig: placeholder}]{%
        \includegraphics[width=0.39\linewidth]{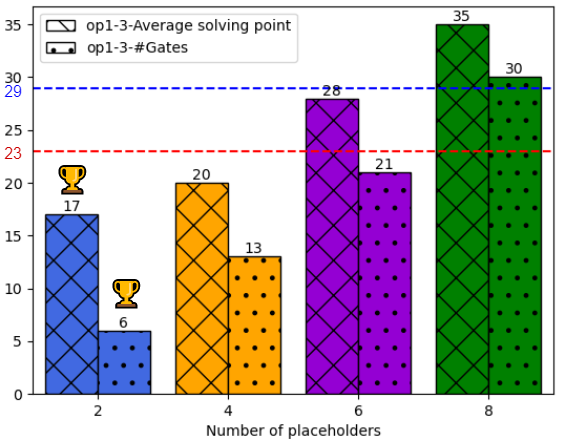}}
    \quad
    \subfloat[Gates and ASP of auto-generated circuit\label{fig: placeholder-gate}]{%
       \includegraphics[width=0.46\linewidth]{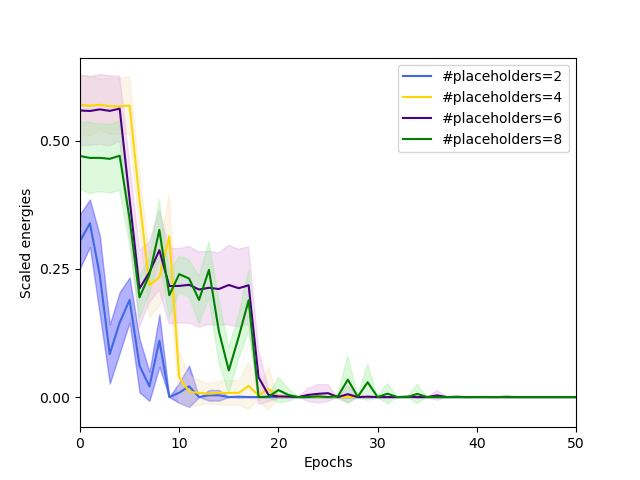}}
    \\
    \subfloat[Comparison of parameterized blocks\label{fig: layer}]{%
       \includegraphics[width=0.39\linewidth]{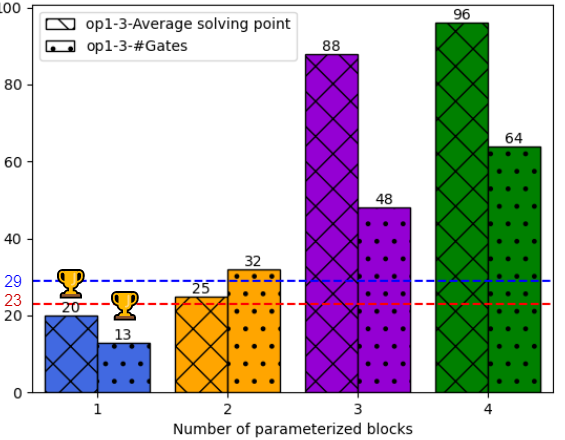}}
   \quad
   \subfloat[Gates and ASP of auto-generated circuit\label{fig: layer-gate}]{%
      \includegraphics[width=0.46\linewidth]{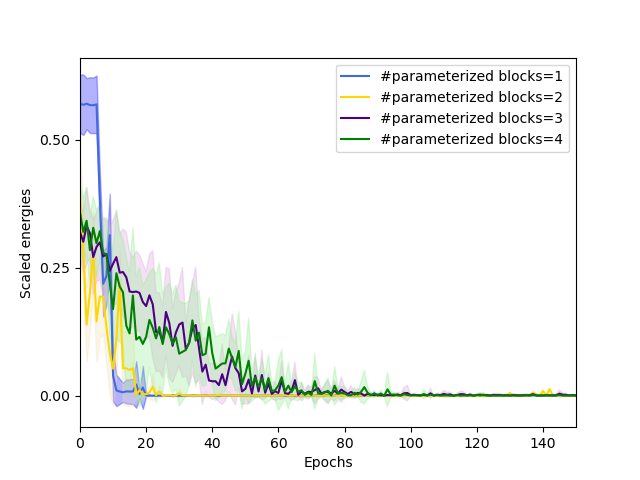}}
  \caption{(a) and (b) illustrate the impact of placeholders when generating circuits with op1-3. (c) and (d) illustrate the impact of parameterized blocks when generating circuits with op1-3. The red lines in (a) and (b) indicate that the baseline circuit has 23 gates, while the ASP has 29 gates.}
  \label{fig:compare}
\end{figure}

The impact of placeholders and parameterized blocks during generating circuits is studied in Fig. \ref{fig:compare}. The evaluation results show that as the number of placeholders and parameterized blocks increases, the number of gates and the depth of the generated circuits also increase, leading to slower convergence when solving a simple JSSP. Deeper circuits require more parameterized gates to be trained, resulting in more complex quantum calculations and, consequently, slower convergence.
\begin{figure}[!ht]
    \centering
    \subfloat[SA-DQAS vs DQAS\label{fig: sa-dqas vs dqas}]{%
        \includegraphics[width=0.47\linewidth]{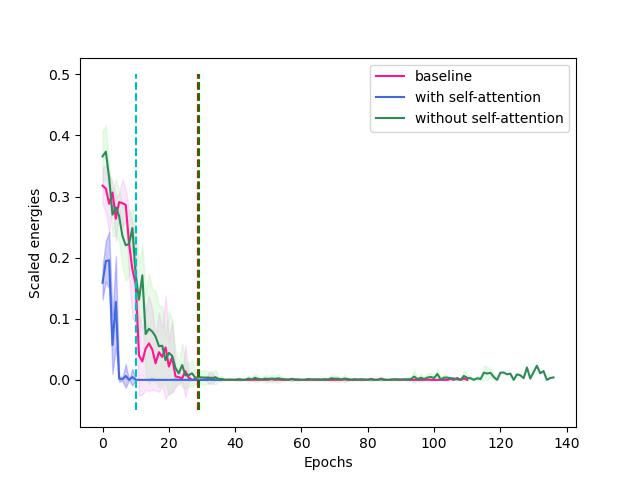}}
    \quad
    \subfloat[Comparison in data\label{fig: comparison in data}]{%
       \includegraphics[width=0.475\linewidth]{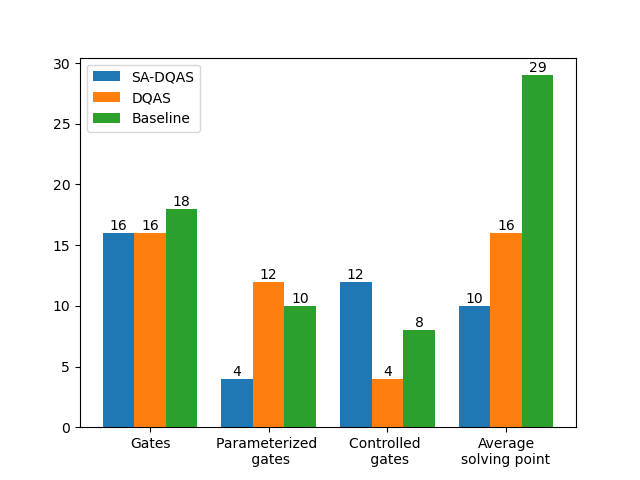}}
    \\
    \subfloat[Circuit by SA-DQAS\label{fig: sa-dqascircuit}]{%
       \includegraphics[width=0.475\linewidth]{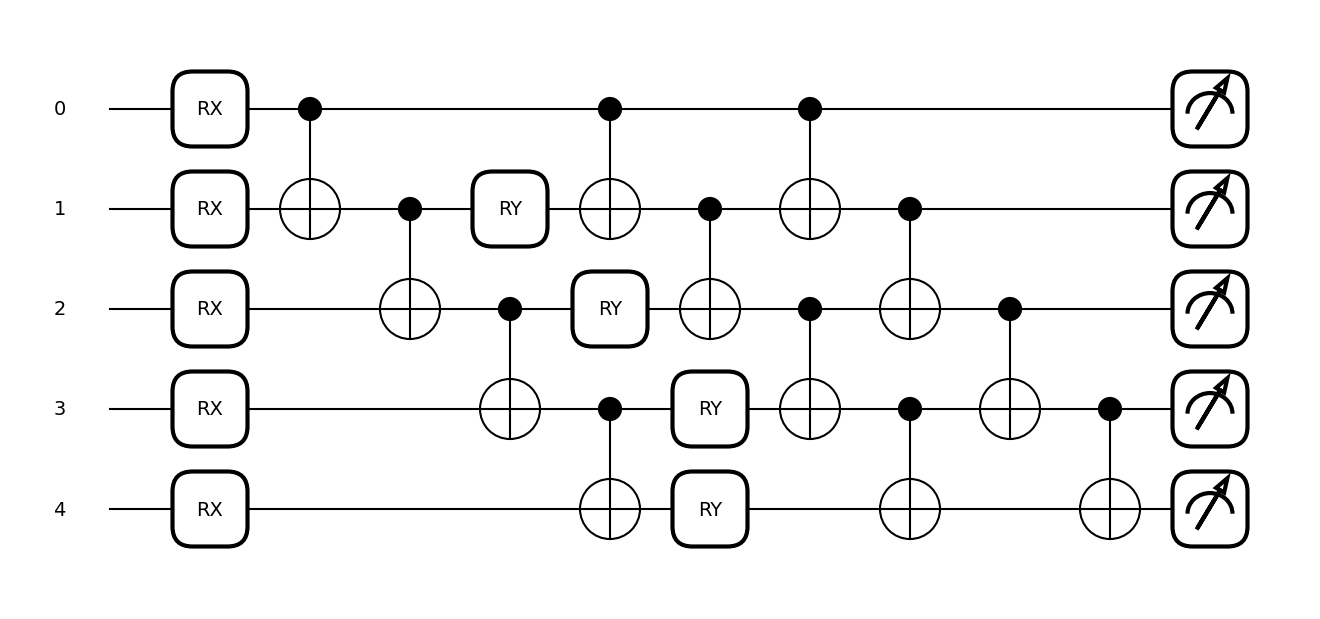}}
   \quad
   \subfloat[Circuit by DQAS\label{fig: dqascircuit}]{%
      \includegraphics[width=0.475\linewidth]{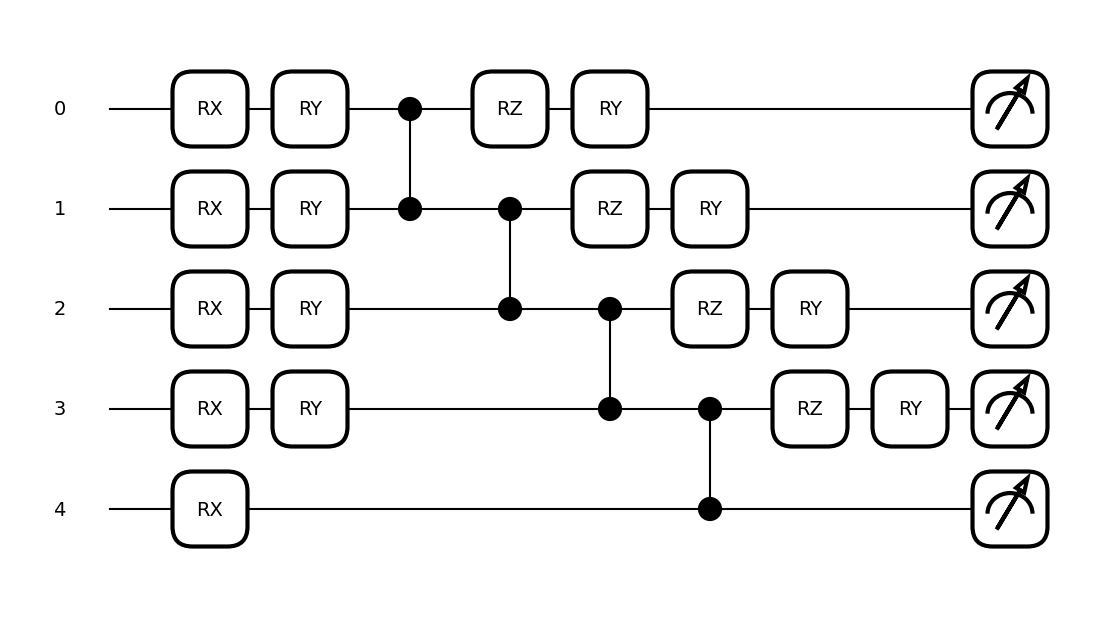}}
    \caption{Comparison of generated circuits.}
  \label{fig:comparedqas}
\end{figure}

We present the comparison of the circuits from op3-3, from which SA-DQAS generates the best-performing circuit. Figure \ref{fig: sa-dqas vs dqas} illustrates that the circuit created with SA-DQAS converges much faster than the baseline and circuit generated with DQAS. Comparing the two circuit architectures in Figures \ref{fig: sa-dqascircuit} and \ref{fig: dqascircuit}, they contain the same number of gates, but more controlled gates appear in \ref{fig: sa-dqascircuit}, meaning that the circuit created with SA-DQAS has fewer trainable parameters, leading to its better performance for JSSP than both the baseline and circuit designed by DQAS. Figure \ref{fig: comparison in data} illustrates the advantages of circuits designed with SA-DQAS mentioned above in the data. 
 \begin{table}[t!]
     \centering
     \caption{Summary of circuits designed with and without self-attention from different operation pools.}
     \begin{threeparttable}
         \begin{tabular}[width=1.0\linewidth]{C{1.1cm}C{0.4cm}C{0.4cm}C{0.4cm}C{0.4cm}C{0.4cm}C{0.4cm}C{0.6cm}C{0.6cm}}
             \hline \multicolumn{1}{C{1.1cm}}{Operation pool} &
                    \multicolumn{2}{C{0.8cm}}{\#Gates$^{***}$} & 
                    \multicolumn{2}{C{0.8cm}}{\#Param. G} &
                    \multicolumn{2}{C{0.8cm}}{\#Con. G}& 
                    \multicolumn{2}{C{1.2cm}}{Avg. solving point$^{**}$} \\
             \hline Baseline & \multicolumn{2}{c}{18} &
                    \multicolumn{2}{c}{10} & \multicolumn{2}{c}{8} &
                    \multicolumn{2}{c}{29}  \\
             \hline    & Y$^{*}$ & N$^{*}$ & Y & N & Y & N & Y & N \\
             \hline op1-1 & 15 & 14 & 15 & 14 & 0 & 0 & \textcolor{green}{25} & 26 \\
                    op1-2 & 13 & 9 & 9 & 9 & 4 & 0 & 18 & 11\\
                    op1-3 & 13 & 19 & 9 & 14 & 4 & 0 & \textcolor{green}{20} & 39\\
                    op1-4 & 14 & 20 & 5 & 15 & 4 & 0 & \textcolor{green}{13} & 38\\
                    op2-1 & 13 & 9 & 9 & 9 & 4 & 0 & \textcolor{green}{13} & 16\\
                    op2-2 & 15 & 16 & 7 & 12 & 8 & 4 & \textcolor{green}{13} & 28\\
                    op2-3 & 16 & 18 & 8 & 18 & 8 & 0 & \textcolor{green}{13} & 40\\
                    op2-4 & 15 & 18 & 10 & 5 & 0 & 8 & 19 & 14\\
                    op3-1 & 15 & 13 & 7 & 9 & 8 & 4 & 18 & 15\\
                    op3-2 & 15 & 13 & 11 & 13 & 4 & 0 & \textcolor{green}{17} & 32\\
                    op3-3 & 16 & 16 & 4 & 12 & 12 & 4 & \textcolor{green}{10} & 16\\
                    op3-4 & 13 & 20 & 5 & 20 & 8 & 0 & \textcolor{green}{14} & 39\\
                    op4-1 & 7 & 12 & 7 & 12 & 0 & 4 & \textcolor{green}{33} & 72\\
                    op4-2 & 12 & 15 & 12 & 15 & 8 & 8 & \textcolor{green}{60} & 108\\
                    op4-3 & 16 & 16 & 16 & 16 & 8 & 8 & \textcolor{green}{89} & 96\\
                    op4-4 & 16 & 17 & 16 & 17 & 16 & 12 & \textcolor{green}{88} & 91\\
             \hline
         \end{tabular}
         \begin{tablenotes}
             \item[*] Y indicates circuits searched by SA-DQAS and N indicates by DQAS
             \item[**] Avg. solving point refers to the number of iterations required to
                      find the minimum energy without subsequent fluctuations.
         \end{tablenotes}
     \end{threeparttable}
     \label{Tab:summaryWith}
 \end{table}

By seeing the information in Table \ref{Tab:summaryWith} in the appendix, we know that nearly all circuits designed with SA-DQAS can find the optimal solution of JSSP faster than those without self-attention. All circuits generated with SA-DQAS have fewer gates than the baseline. When circuits generated with SA-DQAS contain fewer parameterized gates than those with DQAS, they perform better. Nevertheless, there are special cases. The circuit created with SA-DQAS from op1-1 has more parameterized gates but converges faster than with DQAS. They perform fast the same. The circuit created with SA-DQAS from op3-1 performs worse, although it contains less parameterized gates. We can see that it has more controlled gates. This case shows that sometimes controlled gates can reduce training efficiency.

Compared to the structures of circuits generated with different methods, we observe that most circuits created with SA-DQAS contain controlled gates. 
Controlled gates are essential in quantum circuits because they allow us to create interactions and entanglement between quantum qubits, enabling various quantum computation and quantum information processing tasks. A circuit without controlled gates is indistinguishable from a classical circuit. 
Self-attention helps us find better and more meaningful quantum circuits.


\begin{figure}[!ht]
    \centering
    \subfloat[AP-training]{
    \includegraphics[width=0.3\linewidth]{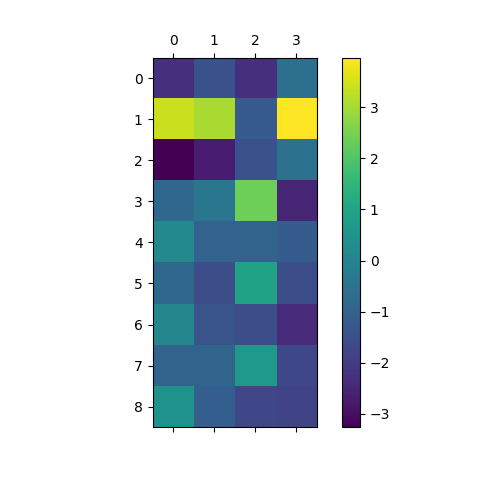}
    \label{fig: ap matrix out}
    }   
    \subfloat[AP-transformer]{
    \includegraphics[width=0.3\linewidth]{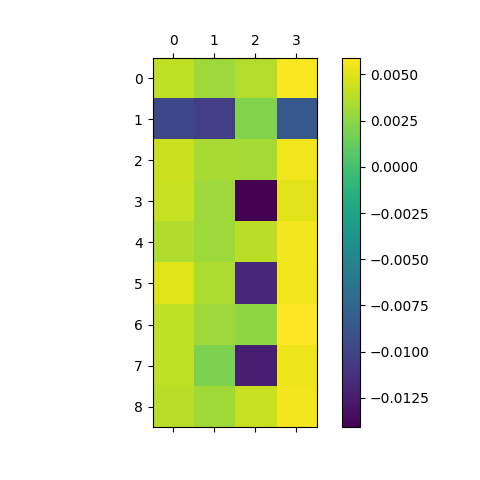}
    \label{fig: ap matrix add}
    }
    \subfloat[AP-end]{
    \includegraphics[width=0.3\linewidth]{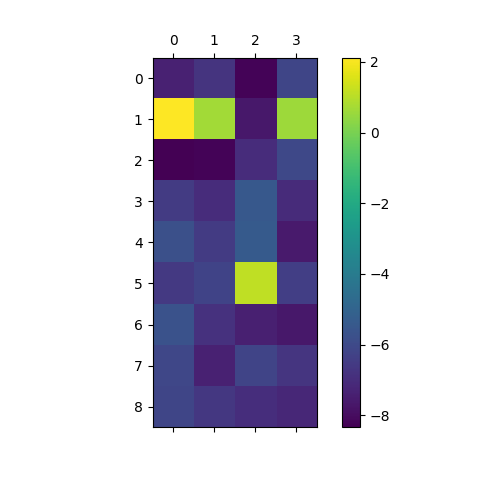}
    \label{fig: ap matrix out end}
    }
    \caption{Illustration of architecture parameters (AP) during training for JSSP problem. In the operation pool there are 9 operation candidates for 4 placeholders, so the architecture parameter matrix here has 9 rows and 4 columns. The subfigure~\ref{fig: ap matrix out} shows the architecture parameter matrix during the training, the lighter the color is, the more likely the operation is to be selected. The subfigure~\ref{fig: ap matrix add} shows the output of transformer encoder which is used to improve the architecture parameters at the same training step. The subfigure~\ref{fig: ap matrix out end} shows the architecture parameters after training.}
  \label{fig:archite para matrix}
\end{figure}

\subsection{Quantum chemistry}
We compared SA-DQAS with different algorithms in terms of minimum energy error of molecules in Table~\ref{Tab:summary-different-algos}. The SA-DQAS can find minimum energy with $10^{-5}$ error that outperforms QCAS~\cite{du2022quantum}, RS or DQAS for molecules $H_2$, $LiH$-$4$ and $LiH$-$6$. But, by comparing with quantumDARTS~\cite{wu2023quantumdarts}, it has similar or slightly lower performance. UCCSD~\cite{romero2018strategies} here gives nearly a theoretical bounds for the minimum energy error.
\begin{table}[t!]
     \centering
     \caption{Experiments on molecules. We solve the Hamiltonian Error of $H_2$, $LiH$-$4$ and $LiH$-$6$ and compare our method to other algorithms. The value indicates the error level compared with ground state energy.}
     \resizebox{0.47\textwidth}{!}{%
     \begin{tabular}{||c | c | c |c||} 
     \hline
     Algorithm & $H_2$ & $LiH$-$4$ & $LiH$-$6$\\
     \hline
     UCCSD & $5.5 \times 10^{-11}$ & $4.0 \times 10^{-5}$ & $4.0 \times 10^{-5}$\\
     quantumDARTS & $4.3 \times 10^{-6}$ & $1.7 \times 10^{-4}$& $2.9 \times 10^{-4}$\\
     DQAS & $3.1 \times 10^{-4}$ & $5.3\times 10^{-4}$& $1.5\times 10^{-3}$\\
     QCAS & $2.2 \times 10^{-2}$ & $8.6\times 10^{-2}$& $7.3\times 10^{-2}$\\
     Randomness & $1.9 \times 10^{-2}$ & $1.3\times 10^{-2}$& $6.2\times 10^{-3}$\\
     \hline
     SA-DQAS & $1.2 \times 10^{-5}$ & $ 3.0\times 10^{-4}$& $ 3.0\times 10^{-4}$\\
     \hline
     \end{tabular}}
     \label{Tab:summary-different-algos}
 \end{table}
\subsection{Fidelity Measurement}
\begin{figure}[!ht]
    \centering
    \subfloat[Ideal circuit]{
    \includegraphics[width=0.35\linewidth]{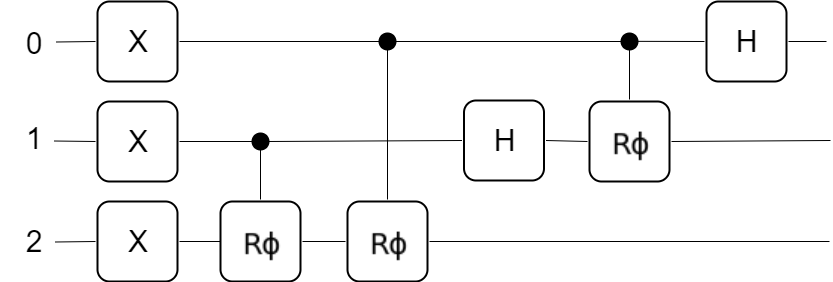}
    \label{fig: idealCircuit}
    }
    \quad
    \subfloat[Circuit from SA-DQAS]{
    \includegraphics[width=0.5\linewidth]{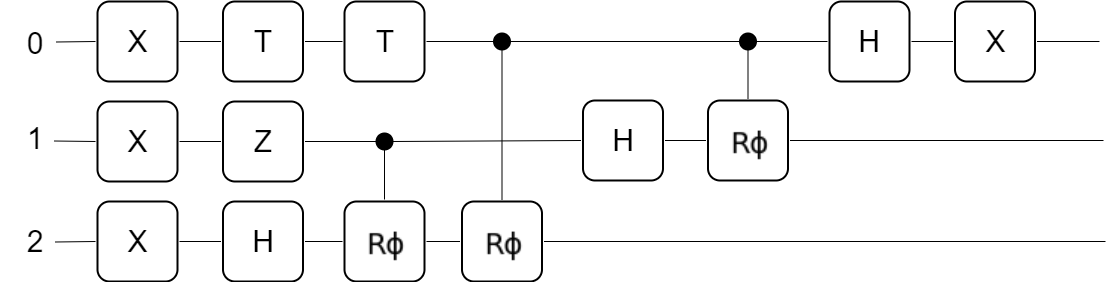}
    \label{fig: bestFidelitySa}
    }
  \caption{Evaluation of newly discovered architectures on a noise-free simulator. The results are averaged over 10 trials with different initial parameters.}
  \label{fig: FidelityCircuits}
\end{figure}

In this section, we test our algorithm for error mitigation and study the fidelity between pre-defined circuits and those generated by SA-DQAS. A previous experiment in~\cite{zhang2022differentiable} demonstrated that DQAS could be used for error mitigation. The pre-defined circuit is a three-qubit \texttt{QFT} circuit as shown in Fig.~\ref{fig: idealCircuit}. During the search process, we set up six placeholders in the circuit—half before and half after the \texttt{QFT} part—following the original experimental setup in~\cite{zhang2022differentiable}. The operation pool used in this experiment is $\mathcal{O}_{f1}$ (shown in Appendix \ref{settings}). The noise model assumes 2\% BitFlip errors between two gates and 20\% BitFlip errors when a qubit is idle. We design circuits using SA-DQAS in an ideal environment and select the one with the highest fidelity to the ideal circuit in a noisy environment. The fidelity between the auto-generated circuit by SA-DQAS (illustrated in Fig. \ref{fig: idealCircuit}) and the ideal circuit is 0.66, which is slightly higher than the fidelity (0.6) between the ideal circuit from DQAS. This result indicates that SA-DQAS outperforms DQAS in error mitigation.

We build three additional noise models based on BitFlip errors. The first model introduces BitFlip errors with varying probabilities at the end of each qubit. The second model adds 2\% BitFlip errors after each operation in non-empty placeholders (where the chosen operation is not "\texttt{E}") of the generated circuit, and a 20\% BitFlip error to each qubit at the end of the circuit. The third model uses the same noise configuration as provided in~\cite{zhang2022differentiable}. We then calculate the fidelity between the ideal circuit and the generated circuits in environments where BitFlip errors occur with different probabilities. 
\begin{table}[!ht]
    \centering
    \caption{A table that records the fidelity between the generated circuit and the ideal circuit under different probabilities of BitFlip noise, with placeholders positioned at various locations. The circuits are automatically generated under an ideal environment (B$_{0.0}$) and with 20\% BitFlip noise (B$_{0.2}$).}
    \label{Tab:summaryBit}
    \begin{threeparttable}
        \begin{tabular}[width=1.0\textwidth]{cccccc}
            \hline Setting & & B$_{0.0}$ & B$_{0.1}$ & B$_{0.2}$ & B$_{0.3}$ \\
            \hline Back & B$_{0.0}$ & 0.94 & 0.79 & 0.68 & 0.57 \\
                        & B$_{0.2}$ & \textcolor{green}{0.97} & \textcolor{green}{0.81} & \textcolor{green}{0.69} & 0.57 \\
            \hline Back and Front & B$_{0.0}$ & \textcolor{green}{0.97} & \textcolor{green}{0.79} & \textcolor{green}{0.68} & 0.57 \\
                        & B$_{0.2}$ & 0.94 & 0.78 & 0.67 & 0.57 \\
            \hline Front & B$_{0.0}$ & 0.99 & 0.77 & 0.65 & 0.54 \\
                        & B$_{0.2}$ & 0.99 & \textcolor{green}{0.79} & \textcolor{green}{0.66} & \textcolor{green}{0.55} \\
            \hline
        \end{tabular}
    \end{threeparttable}
\end{table}

From the results in Table \ref{Tab:summaryBit}, we observe that as the probability of BitFlip occurrence increases, the average fidelity decreases. When placeholders are positioned entirely before or after the \texttt{QFT} part, circuits generated in a noisy environment exhibit higher average fidelity than those created in an ideal environment. However, when the \texttt{QFT} part is in the middle of the circuit, the situation is reversed. Circuits generated in noisy environments tend to be more stable and reliable, with better structures that are less affected by BitFlip noise, making them more noise-resilient. Additionally, the noise resistance of the generated circuits is also influenced by their structural design. In Appendix\ref{fidelity}, we also show other noise models experiments based on PhaseDamping, AmplitudeDamping, DepolarizingChannel, and other settings.


\subsection{Real-Device Evaluation and Scalability via MaxCut}

While our previous experiments used quantum simulators and employed smaller circuit sizes, a key question still stands: \textit{How do searched architectures scale to real quantum hardware, and can they scale to bigger problem instances?} To answer this, we have a new, complementary experiment, testing the MaxCut problem, a standard benchmark in quantum combinatorial optimization.

In order to verify searched circuit architectures for scale, we use a stacking method. Precisely, we train circuit block on tiny graphs. We generate each circuit by block which shares the same architecture and represents each edge between two nodes. Subsequently, we examine their performances when stacking these blocks to create deeper circuits for larger graphs. This replicates real-world scenarios in which deep circuits are built from transferable units that have been trained in smaller subproblems.

We examine two graph families, complete bipartite graphs and random graphs, in both $n=8$ and $n=16$ nodes. The experiments ran with real quantum hardware \texttt{ibmq\_sherbrooke} with 10,000 shots per circuit.

\begin{table*}[t]
\centering
\caption{Performance of Stacked Circuit Architectures on Real Quantum Devices (MaxCut)}
\label{tab:real_device_maxcut}
\begin{tabular}{|l|c|c|l|c|l|}
\hline
\textbf{Graph Type} & \textbf{\#Qubits} & \textbf{Architecture} & \textbf{Cost (Target)} & \textbf{Shots} & \textbf{Device} \\
\hline
Complete Bipartite ($n=8$)  & 8  & cnot-ry-cnot & 16 (16) & 10,000 & \texttt{ibmq\_sherbrooke} \\
Complete Bipartite ($n=16$) & 16 & cnot-ry-cnot & 64 (64) & 10,000 & \texttt{ibmq\_sherbrooke} \\
Random ($n=8$)              & 8  & cnot-ry-cnot & 12 (12) & 10,000 & \texttt{ibmq\_sherbrooke} \\
Random ($n=16$)             & 16 & cnot-ry-cnot & 20 (20) & 10,000 & \texttt{ibmq\_sherbrooke} \\
\hline
\end{tabular}
\end{table*}

\begin{figure}[t]
\centering
\begin{tabular}{cc}
\includegraphics[width=0.22\textwidth]{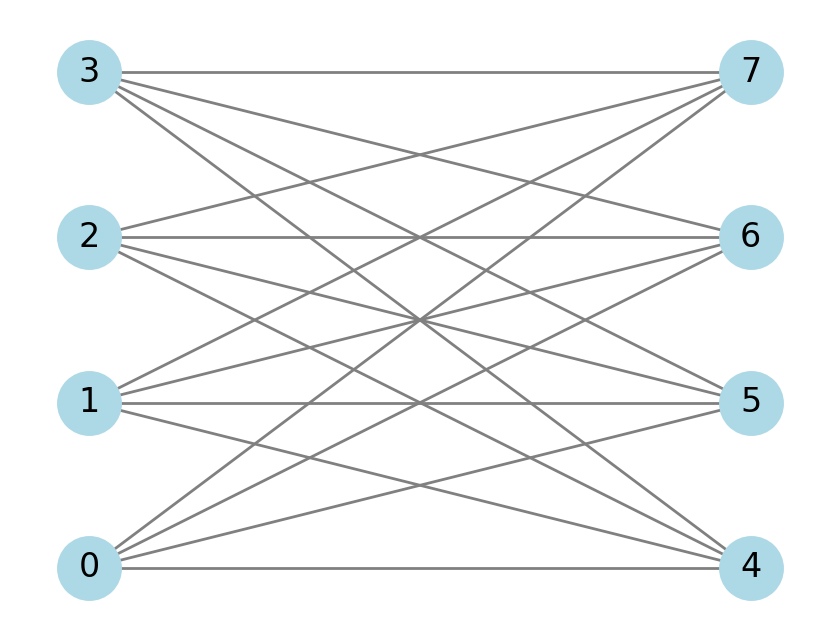} &
\includegraphics[width=0.22\textwidth]{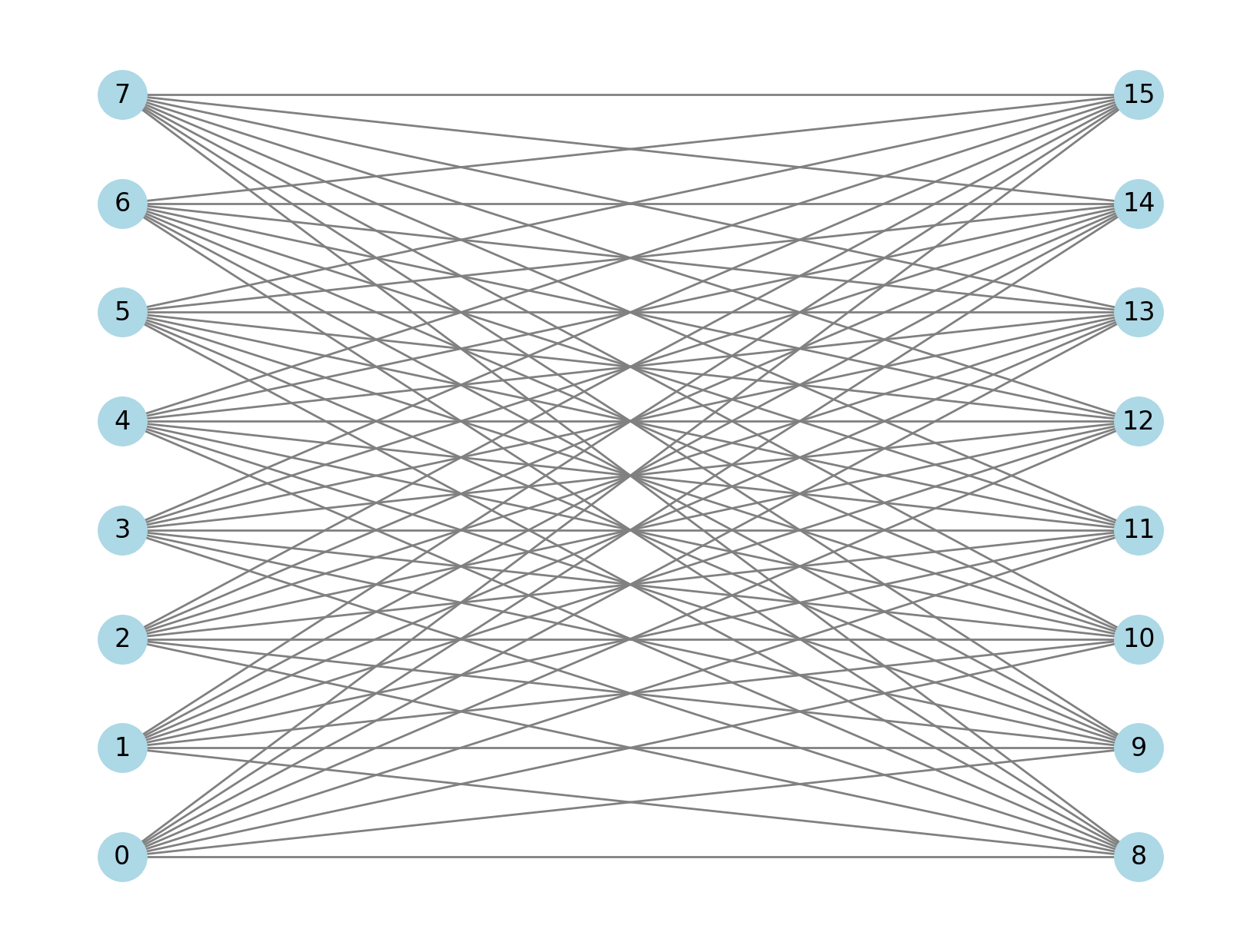} \\
(a) Bipartite (8 nodes) & (b) Bipartite (16 nodes) \\
\includegraphics[width=0.22\textwidth]{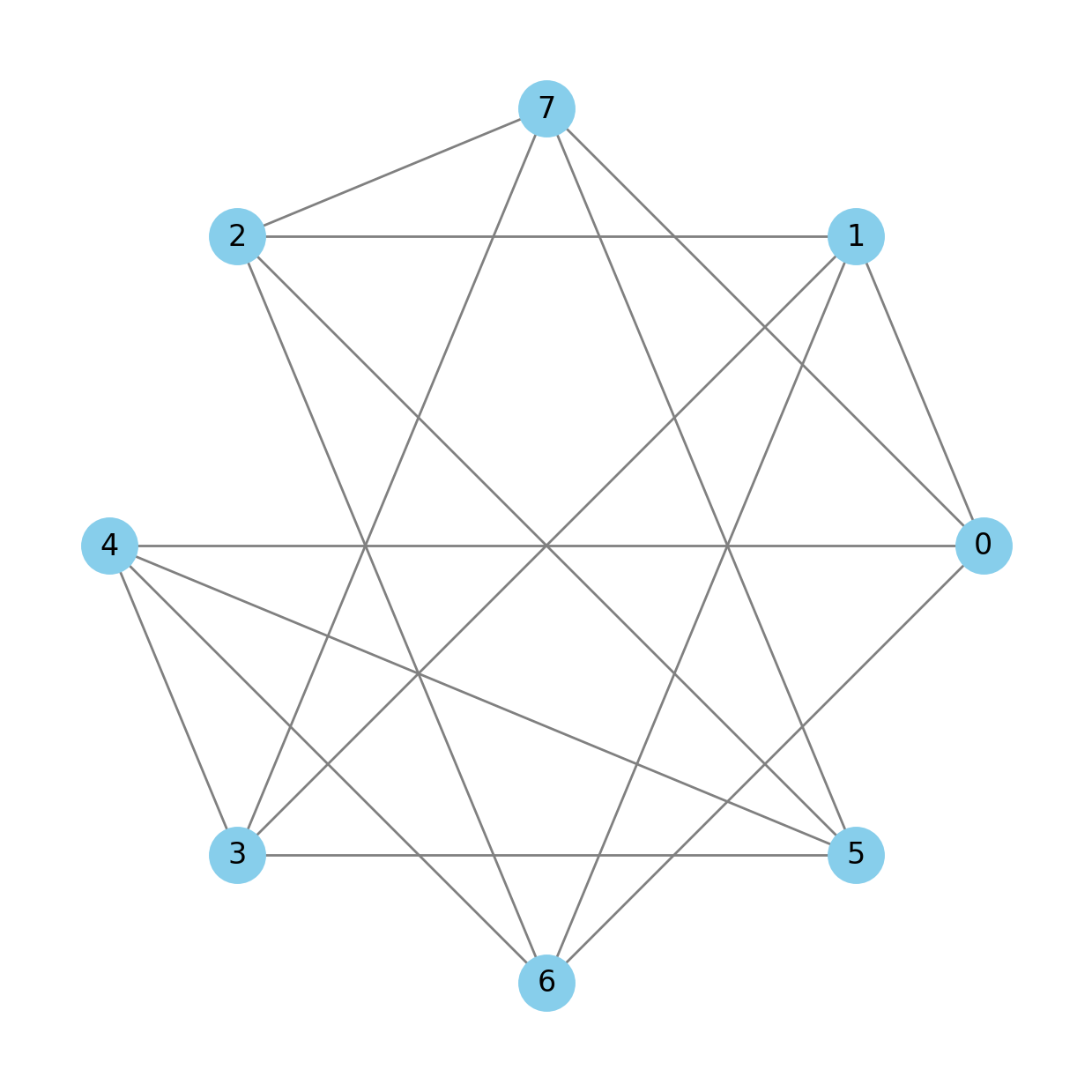} &
\includegraphics[width=0.22\textwidth]{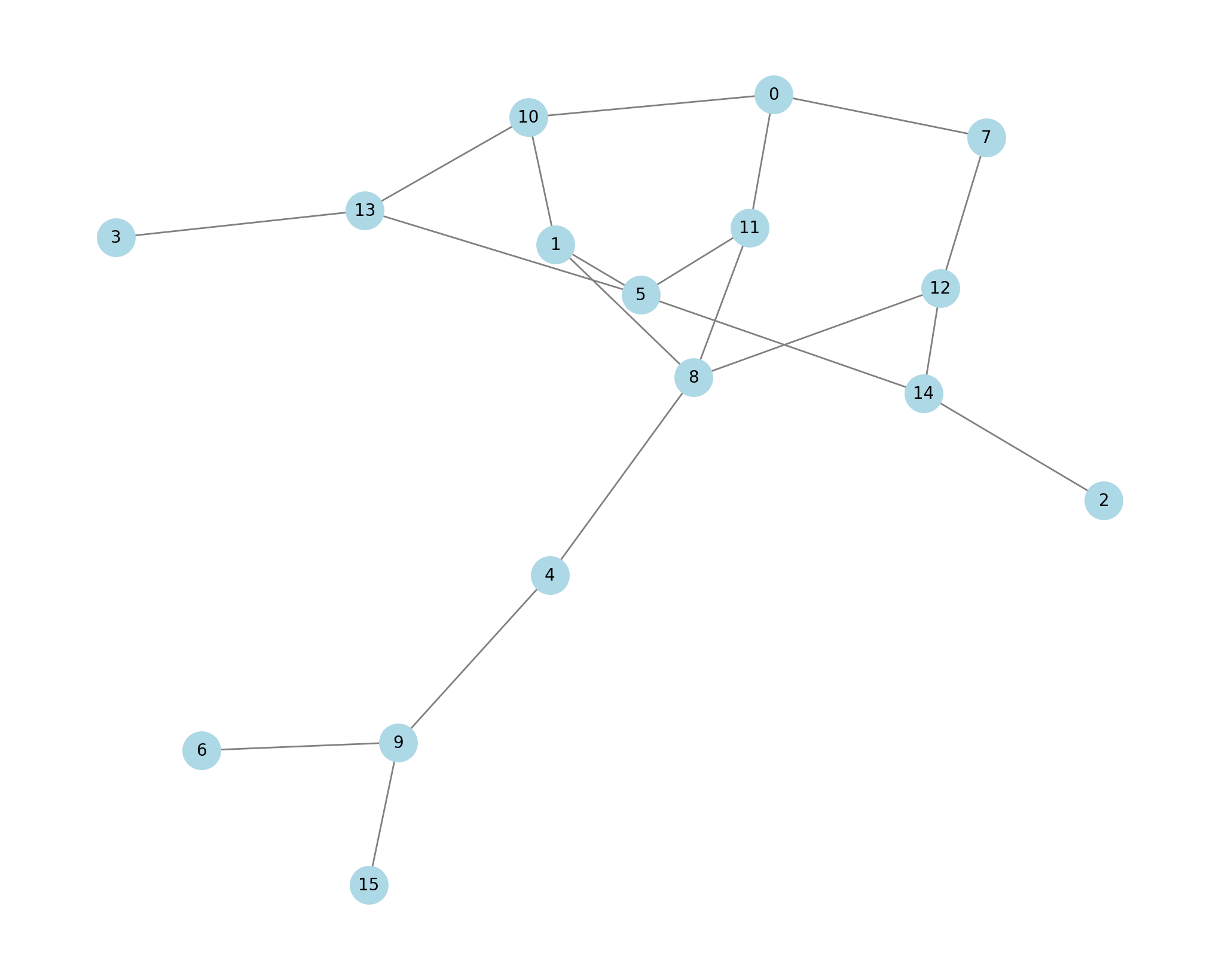} \\
(c) Random (8 nodes) & (d) Random (16 nodes)
\end{tabular}
\caption{Graphs used in real-device MaxCut experiments. Complete bipartite and random graphs of different sizes are used to evaluate the scalability and generalization performance of stacked circuit architectures.}
\label{fig:maxcut_graphs}
\end{figure}

The values in Table~\ref{tab:real_device_maxcut} show that stacked architecture achieves optimal or nearly optimal values of cost even on real hardware as well as on bigger graphs. This showcases that not only do searched architectures generalize over different instances of a problem, but they also scale moderately when applied to bigger circuits by repeated replication. Significantly, it proves the deployability of our NAS-learned circuits to near-term quantum hardware.

\section{Conclusion}
We have introduced SA-DQAS, which extends Differentiable Quantum Architecture Search (DQAS) to use self-attention to learn interrelations among candidate operations in placeholder positions. This global viewpoint produces more stable, high-performing quantum architectures. Our approach, which we evaluate over a variety of quantum problems including MaxCut, JSSP, quantum chemistry, and fidelity optimization, consistently outperforms DQAS as well as other baselines in terms of convergence speed, quality of the solution, and robustness to noise. Specifically, we introduce real-device evaluation on IBM's hardware via MaxCut, where architecture found in small graph training is stacked and used to tackle large-scale graphs. This scale test demonstrates SA-DQAS's ability to create transferable, hardware-efficient blocks of circuits, pushing its practical applications.

\nocite{openai2024chatgpt}

\begin{acks}
The project of this conference paper is based on was supported with funds from the
German Federal Ministry of Education and Research in the funding program
Quantum Reinforcement Learning for industrial Applications (QLindA) and the Federal Ministry for Economic Affairs and Climate Action
in the funding program Quantum-Classical Hybrid Optimization Algorithms for Logistics
and Production Line Management (QCHALLenge)
- under project number 13N15644 and 01MQ22008B.
AI assistance was used during the preparation of this manuscript for tasks such as drafting, paraphrasing, and grammar suggestions, using ChatGPT (OpenAI, 2024). The authors have reviewed and revised all content for accuracy and originality. Due to the iterative nature of writing, we cannot precisely identify individual sentences influenced by AI assistance.
\end{acks}

\bibliographystyle{ACM-Reference-Format}
\bibliography{sample-base}


\begin{thebibliography}{35}


\ifx \showCODEN    \undefined \def \showCODEN     #1{\unskip}     \fi
\ifx \showISBNx    \undefined \def \showISBNx     #1{\unskip}     \fi
\ifx \showISBNxiii \undefined \def \showISBNxiii  #1{\unskip}     \fi
\ifx \showISSN     \undefined \def \showISSN      #1{\unskip}     \fi
\ifx \showLCCN     \undefined \def \showLCCN      #1{\unskip}     \fi
\ifx \shownote     \undefined \def \shownote      #1{#1}          \fi
\ifx \showarticletitle \undefined \def \showarticletitle #1{#1}   \fi
\ifx \showURL      \undefined \def \showURL       {\relax}        \fi
\providecommand\bibfield[2]{#2}
\providecommand\bibinfo[2]{#2}
\providecommand\natexlab[1]{#1}
\providecommand\showeprint[2][]{arXiv:#2}

\bibitem[Alam et~al\mbox{.}(2021)]%
        {alam2021quantum}
\bibfield{author}{\bibinfo{person}{Mahabubul Alam}, \bibinfo{person}{Satwik Kundu}, \bibinfo{person}{Rasit~Onur Topaloglu}, {and} \bibinfo{person}{Swaroop Ghosh}.} \bibinfo{year}{2021}\natexlab{}.
\newblock \showarticletitle{Quantum-Classical Hybrid Machine Learning for Image Classification (ICCAD Special Session Paper)}.
\newblock \bibinfo{journal}{\emph{arXiv preprint arXiv:2109.02862}} (\bibinfo{year}{2021}).
\newblock


\bibitem[Amaro et~al\mbox{.}(2022)]%
        {amaro2022case}
\bibfield{author}{\bibinfo{person}{David Amaro}, \bibinfo{person}{Matthias Rosenkranz}, \bibinfo{person}{Nathan Fitzpatrick}, \bibinfo{person}{Koji Hirano}, {and} \bibinfo{person}{Mattia Fiorentini}.} \bibinfo{year}{2022}\natexlab{}.
\newblock \showarticletitle{A case study of variational quantum algorithms for a job shop scheduling problem}.
\newblock \bibinfo{journal}{\emph{EPJ Quantum Technology}} \bibinfo{volume}{9}, \bibinfo{number}{1} (\bibinfo{year}{2022}), \bibinfo{pages}{5}.
\newblock


\bibitem[Bahdanau et~al\mbox{.}(2015)]%
        {bahdanau2014neural}
\bibfield{author}{\bibinfo{person}{Dzmitry Bahdanau}, \bibinfo{person}{Kyunghyun Cho}, {and} \bibinfo{person}{Yoshua Bengio}.} \bibinfo{year}{2015}\natexlab{}.
\newblock \showarticletitle{Neural Machine Translation by Jointly Learning to Align and Translate}. In \bibinfo{booktitle}{\emph{International Conference on Learning Representations (ICLR)}}. \bibinfo{publisher}{ICLR}, \bibinfo{address}{San Diego, CA, USA}.
\newblock
\newblock
\shownote{No page numbers (arXiv preprint arXiv:1409.0473)}.


\bibitem[Carugno et~al\mbox{.}(2022)]%
        {carugno2022evaluating}
\bibfield{author}{\bibinfo{person}{Costantino Carugno}, \bibinfo{person}{Maurizio Ferrari~Dacrema}, {and} \bibinfo{person}{Paolo Cremonesi}.} \bibinfo{year}{2022}\natexlab{}.
\newblock \showarticletitle{Evaluating the job shop scheduling problem on a D-wave quantum annealer}.
\newblock \bibinfo{journal}{\emph{Scientific Reports}} \bibinfo{volume}{12}, \bibinfo{number}{1} (\bibinfo{year}{2022}), \bibinfo{pages}{6539}.
\newblock


\bibitem[Cerezo et~al\mbox{.}(2021)]%
        {cerezo2021variational}
\bibfield{author}{\bibinfo{person}{Marco Cerezo}, \bibinfo{person}{Andrew Arrasmith}, \bibinfo{person}{Ryan Babbush}, \bibinfo{person}{Simon~C Benjamin}, \bibinfo{person}{Suguru Endo}, \bibinfo{person}{Keisuke Fujii}, \bibinfo{person}{Jarrod~R McClean}, \bibinfo{person}{Kosuke Mitarai}, \bibinfo{person}{Xiao Yuan}, \bibinfo{person}{Lukasz Cincio}, {et~al\mbox{.}}} \bibinfo{year}{2021}\natexlab{}.
\newblock \showarticletitle{Variational quantum algorithms}.
\newblock \bibinfo{journal}{\emph{Nature Reviews Physics}} \bibinfo{volume}{3}, \bibinfo{number}{9} (\bibinfo{year}{2021}), \bibinfo{pages}{625--644}.
\newblock


\bibitem[Chen and Kuo(2025)]%
        {chen2025quantum}
\bibfield{author}{\bibinfo{person}{Chi-Sheng Chen} {and} \bibinfo{person}{En-Jui Kuo}.} \bibinfo{year}{2025}\natexlab{}.
\newblock \showarticletitle{Quantum Adaptive Self-Attention for Quantum Transformer Models}.
\newblock \bibinfo{journal}{\emph{arXiv preprint arXiv:2504.05336}} (\bibinfo{year}{2025}).
\newblock


\bibitem[Chen et~al\mbox{.}(2024)]%
        {chen2024quantumqtrain}
\bibfield{author}{\bibinfo{person}{Kuan-Cheng Chen}, \bibinfo{person}{Samuel Yen-Chi Chen}, \bibinfo{person}{Chen-Yu Liu}, {and} \bibinfo{person}{Kin~K Leung}.} \bibinfo{year}{2024}\natexlab{}.
\newblock \showarticletitle{Quantum-Train-Based Distributed Multi-Agent Reinforcement Learning}.
\newblock \bibinfo{journal}{\emph{arXiv preprint arXiv:2412.08845}} (\bibinfo{year}{2024}).
\newblock


\bibitem[Chen(2024a)]%
        {chen2024differentiable}
\bibfield{author}{\bibinfo{person}{Samuel Yen-Chi Chen}.} \bibinfo{year}{2024}\natexlab{a}.
\newblock \showarticletitle{Differentiable quantum architecture search in asynchronous quantum reinforcement learning}. In \bibinfo{booktitle}{\emph{2024 IEEE International Conference on Quantum Computing and Engineering (QCE)}}, Vol.~\bibinfo{volume}{1}. \bibinfo{publisher}{IEEE}, \bibinfo{address}{Piscataway, NJ, USA}, \bibinfo{pages}{1516--1524}.
\newblock


\bibitem[Chen(2024b)]%
        {chen2024evolutionary}
\bibfield{author}{\bibinfo{person}{Samuel Yen-Chi Chen}.} \bibinfo{year}{2024}\natexlab{b}.
\newblock \bibinfo{title}{Differentiable Quantum Architecture Search in Asynchronous Quantum Reinforcement Learning}.
\newblock
\showeprint[arxiv]{2407.18202}~[cs.LG]
\newblock
\shownote{Preprint, arXiv:2407.18202}.


\bibitem[Chen et~al\mbox{.}(2025)]%
        {chen2025differentiable}
\bibfield{author}{\bibinfo{person}{Samuel Yen-Chi Chen}, \bibinfo{person}{Chen-Yu Liu}, \bibinfo{person}{Kuan-Cheng Chen}, \bibinfo{person}{Wei-Jia Huang}, \bibinfo{person}{Yen-Jui Chang}, {and} \bibinfo{person}{Wei-Hao Huang}.} \bibinfo{year}{2025}\natexlab{}.
\newblock \showarticletitle{Differentiable Quantum Architecture Search in Quantum-Enhanced Neural Network Parameter Generation}.
\newblock \bibinfo{journal}{\emph{arXiv preprint arXiv:2505.09653}} (\bibinfo{year}{2025}).
\newblock


\bibitem[Dai et~al\mbox{.}(2024)]%
        {dai2024quantum}
\bibfield{author}{\bibinfo{person}{Xin Dai}, \bibinfo{person}{Tzu-Chieh Wei}, \bibinfo{person}{Shinjae Yoo}, {and} \bibinfo{person}{Samuel Yen-Chi Chen}.} \bibinfo{year}{2024}\natexlab{}.
\newblock \showarticletitle{Quantum Machine Learning Architecture Search via Deep Reinforcement Learning}.
\newblock \bibinfo{journal}{\emph{arXiv preprint arXiv:2407.20147}} (\bibinfo{year}{2024}).
\newblock


\bibitem[Devlin et~al\mbox{.}(2018)]%
        {devlin2018bert}
\bibfield{author}{\bibinfo{person}{Jacob Devlin}, \bibinfo{person}{Ming-Wei Chang}, \bibinfo{person}{Kenton Lee}, {and} \bibinfo{person}{Kristina Toutanova}.} \bibinfo{year}{2018}\natexlab{}.
\newblock \showarticletitle{Bert: Pre-training of deep bidirectional transformers for language understanding}.
\newblock \bibinfo{journal}{\emph{arXiv preprint arXiv:1810.04805}} (\bibinfo{year}{2018}).
\newblock


\bibitem[Ding and Spector(2022)]%
        {ding2022evolutionary}
\bibfield{author}{\bibinfo{person}{Li Ding} {and} \bibinfo{person}{Lee Spector}.} \bibinfo{year}{2022}\natexlab{}.
\newblock \showarticletitle{Evolutionary quantum architecture search for parametrized quantum circuits}. In \bibinfo{booktitle}{\emph{Proceedings of the Genetic and Evolutionary Computation Conference Companion}}. \bibinfo{publisher}{Association for Computing Machinery}, \bibinfo{address}{New York, NY, USA}, \bibinfo{pages}{2190--2195}.
\newblock


\bibitem[Ding and Spector(2023)]%
        {ding2023multi}
\bibfield{author}{\bibinfo{person}{Li Ding} {and} \bibinfo{person}{Lee Spector}.} \bibinfo{year}{2023}\natexlab{}.
\newblock \showarticletitle{Multi-Objective Evolutionary Architecture Search for Parameterized Quantum Circuits}.
\newblock \bibinfo{journal}{\emph{Entropy}} \bibinfo{volume}{25}, \bibinfo{number}{1} (\bibinfo{year}{2023}), \bibinfo{pages}{93}.
\newblock


\bibitem[Du et~al\mbox{.}(2022)]%
        {du2022quantum}
\bibfield{author}{\bibinfo{person}{Yuxuan Du}, \bibinfo{person}{Tao Huang}, \bibinfo{person}{Shan You}, \bibinfo{person}{Min-Hsiu Hsieh}, {and} \bibinfo{person}{Dacheng Tao}.} \bibinfo{year}{2022}\natexlab{}.
\newblock \showarticletitle{Quantum circuit architecture search for variational quantum algorithms}.
\newblock \bibinfo{journal}{\emph{npj Quantum Information}} \bibinfo{volume}{8}, \bibinfo{number}{1} (\bibinfo{year}{2022}), \bibinfo{pages}{1--8}.
\newblock


\bibitem[F{\"o}sel et~al\mbox{.}(2021)]%
        {fosel2021quantum}
\bibfield{author}{\bibinfo{person}{Thomas F{\"o}sel}, \bibinfo{person}{Murphy~Yuezhen Niu}, \bibinfo{person}{Florian Marquardt}, {and} \bibinfo{person}{Li Li}.} \bibinfo{year}{2021}\natexlab{}.
\newblock \showarticletitle{Quantum circuit optimization with deep reinforcement learning}.
\newblock \bibinfo{journal}{\emph{arXiv preprint arXiv:2103.07585}} (\bibinfo{year}{2021}).
\newblock


\bibitem[Gehring et~al\mbox{.}(2017)]%
        {gehring2017convolutional}
\bibfield{author}{\bibinfo{person}{Jonas Gehring}, \bibinfo{person}{Michael Auli}, \bibinfo{person}{David Grangier}, \bibinfo{person}{Denis Yarats}, {and} \bibinfo{person}{Yann~N Dauphin}.} \bibinfo{year}{2017}\natexlab{}.
\newblock \showarticletitle{Convolutional sequence to sequence learning}. In \bibinfo{booktitle}{\emph{Proceedings of the 34th International Conference on Machine Learning}}. \bibinfo{publisher}{PMLR}, \bibinfo{address}{Sydney, Australia}, \bibinfo{pages}{1243--1252}.
\newblock


\bibitem[Giovagnoli et~al\mbox{.}(2023)]%
        {giovagnoli2023qneat}
\bibfield{author}{\bibinfo{person}{Alessandro Giovagnoli}, \bibinfo{person}{Volker Tresp}, \bibinfo{person}{Yunpu Ma}, {and} \bibinfo{person}{Matthias Schubert}.} \bibinfo{year}{2023}\natexlab{}.
\newblock \showarticletitle{Qneat: Natural evolution of variational quantum circuit architecture}. In \bibinfo{booktitle}{\emph{Proceedings of the Companion Conference on Genetic and Evolutionary Computation}}. \bibinfo{pages}{647--650}.
\newblock


\bibitem[He et~al\mbox{.}(2023)]%
        {he2023gsqas}
\bibfield{author}{\bibinfo{person}{Zhimin He}, \bibinfo{person}{Maijie Deng}, \bibinfo{person}{Shenggen Zheng}, \bibinfo{person}{Lvzhou Li}, {and} \bibinfo{person}{Haozhen Situ}.} \bibinfo{year}{2023}\natexlab{}.
\newblock \showarticletitle{GSQAS: Graph Self-supervised Quantum Architecture Search}.
\newblock \bibinfo{journal}{\emph{arXiv preprint arXiv:2303.12381}} (\bibinfo{year}{2023}).
\newblock


\bibitem[Kim et~al\mbox{.}(2025)]%
        {kim2025quantum}
\bibfield{author}{\bibinfo{person}{Gyu~Seon Kim}, \bibinfo{person}{Sungjoon Lee}, \bibinfo{person}{In-Sop Cho}, \bibinfo{person}{Soohyun Park}, {and} \bibinfo{person}{Joongheon Kim}.} \bibinfo{year}{2025}\natexlab{}.
\newblock \showarticletitle{Quantum Reinforcement Learning for Lightweight LEO Satellite Routing}.
\newblock \bibinfo{journal}{\emph{IEEE Internet of Things Journal}} (\bibinfo{year}{2025}).
\newblock


\bibitem[Kuo et~al\mbox{.}(2021)]%
        {kuo2021quantum}
\bibfield{author}{\bibinfo{person}{En-Jui Kuo}, \bibinfo{person}{Yao-Lung~L Fang}, {and} \bibinfo{person}{Samuel Yen-Chi Chen}.} \bibinfo{year}{2021}\natexlab{}.
\newblock \showarticletitle{Quantum architecture search via deep reinforcement learning}.
\newblock \bibinfo{journal}{\emph{arXiv preprint arXiv:2104.07715}} (\bibinfo{year}{2021}).
\newblock


\bibitem[Ma et~al\mbox{.}(2019)]%
        {ma2019variational}
\bibfield{author}{\bibinfo{person}{Yunpu Ma}, \bibinfo{person}{Volker Tresp}, \bibinfo{person}{Liming Zhao}, {and} \bibinfo{person}{Yuyi Wang}.} \bibinfo{year}{2019}\natexlab{}.
\newblock \showarticletitle{Variational quantum circuit model for knowledge graph embedding}.
\newblock \bibinfo{journal}{\emph{Advanced Quantum Technologies}} \bibinfo{volume}{2}, \bibinfo{number}{7-8} (\bibinfo{year}{2019}), \bibinfo{pages}{1800078}.
\newblock


\bibitem[Meng et~al\mbox{.}(2021)]%
        {meng2021quantum}
\bibfield{author}{\bibinfo{person}{Fan-Xu Meng}, \bibinfo{person}{Ze-Tong Li}, \bibinfo{person}{Xu-Tao Yu}, {and} \bibinfo{person}{Zai-Chen Zhang}.} \bibinfo{year}{2021}\natexlab{}.
\newblock \showarticletitle{Quantum Circuit Architecture Optimization for Variational Quantum Eigensolver via Monto Carlo Tree Search}.
\newblock \bibinfo{journal}{\emph{IEEE Transactions on Quantum Engineering}}  \bibinfo{volume}{2} (\bibinfo{year}{2021}), \bibinfo{pages}{1--10}.
\newblock


\bibitem[{OpenAI}(2024)]%
        {openai2024chatgpt}
\bibfield{author}{\bibinfo{person}{{OpenAI}}.} \bibinfo{year}{2024}\natexlab{}.
\newblock \bibinfo{title}{ChatGPT}.
\newblock \bibinfo{howpublished}{\url{https://openai.com/chatgpt}}.
\newblock
\newblock
\shownote{Accessed: May 2025}.


\bibitem[Romero et~al\mbox{.}(2018)]%
        {romero2018strategies}
\bibfield{author}{\bibinfo{person}{Jonathan Romero}, \bibinfo{person}{Ryan Babbush}, \bibinfo{person}{Jarrod~R McClean}, \bibinfo{person}{Cornelius Hempel}, \bibinfo{person}{Peter~J Love}, {and} \bibinfo{person}{Al{\'a}n Aspuru-Guzik}.} \bibinfo{year}{2018}\natexlab{}.
\newblock \showarticletitle{Strategies for quantum computing molecular energies using the unitary coupled cluster ansatz}.
\newblock \bibinfo{journal}{\emph{Quantum Science and Technology}} \bibinfo{volume}{4}, \bibinfo{number}{1} (\bibinfo{year}{2018}), \bibinfo{pages}{014008}.
\newblock


\bibitem[Shaw et~al\mbox{.}(2018)]%
        {shaw2018self}
\bibfield{author}{\bibinfo{person}{Peter Shaw}, \bibinfo{person}{Jakob Uszkoreit}, {and} \bibinfo{person}{Ashish Vaswani}.} \bibinfo{year}{2018}\natexlab{}.
\newblock \showarticletitle{Self-attention with relative position representations}.
\newblock \bibinfo{journal}{\emph{arXiv preprint arXiv:1803.02155}} (\bibinfo{year}{2018}).
\newblock


\bibitem[Sun et~al\mbox{.}(2024)]%
        {sun2024quantum}
\bibfield{author}{\bibinfo{person}{Yize Sun}, \bibinfo{person}{Zixin Wu}, \bibinfo{person}{Yunpu Ma}, {and} \bibinfo{person}{Volker Tresp}.} \bibinfo{year}{2024}\natexlab{}.
\newblock \showarticletitle{Quantum Architecture Search with Unsupervised Representation Learning}.
\newblock \bibinfo{journal}{\emph{arXiv preprint arXiv:2401.11576}} (\bibinfo{year}{2024}).
\newblock


\bibitem[Vaswani et~al\mbox{.}(2017)]%
        {vaswani2017attention}
\bibfield{author}{\bibinfo{person}{Ashish Vaswani}, \bibinfo{person}{Noam Shazeer}, \bibinfo{person}{Niki Parmar}, \bibinfo{person}{Jakob Uszkoreit}, \bibinfo{person}{Llion Jones}, \bibinfo{person}{Aidan~N Gomez}, \bibinfo{person}{{\L}ukasz Kaiser}, {and} \bibinfo{person}{Illia Polosukhin}.} \bibinfo{year}{2017}\natexlab{}.
\newblock \showarticletitle{Attention is all you need}.
\newblock \bibinfo{journal}{\emph{Advances in neural information processing systems}}  \bibinfo{volume}{30} (\bibinfo{year}{2017}).
\newblock


\bibitem[Venturelli et~al\mbox{.}(2016)]%
        {venturelli2016job}
\bibfield{author}{\bibinfo{person}{Davide Venturelli}, \bibinfo{person}{D Marchand}, {and} \bibinfo{person}{Galo Rojo}.} \bibinfo{year}{2016}\natexlab{}.
\newblock \showarticletitle{Job shop scheduling solver based on quantum annealing}. In \bibinfo{booktitle}{\emph{Proc. of ICAPS-16 Workshop on Constraint Satisfaction Techniques for Planning and Scheduling (COPLAS)}}. \bibinfo{publisher}{UNKNOWN}, \bibinfo{address}{UNKNOWN}, \bibinfo{pages}{25--34}.
\newblock


\bibitem[Wallman and Emerson(2016)]%
        {wallman2016noise}
\bibfield{author}{\bibinfo{person}{Joel~J Wallman} {and} \bibinfo{person}{Joseph Emerson}.} \bibinfo{year}{2016}\natexlab{}.
\newblock \showarticletitle{Noise tailoring for scalable quantum computation via randomized compiling}.
\newblock \bibinfo{journal}{\emph{Physical Review A}} \bibinfo{volume}{94}, \bibinfo{number}{5} (\bibinfo{year}{2016}), \bibinfo{pages}{052325}.
\newblock


\bibitem[Wu et~al\mbox{.}(2023)]%
        {wu2023quantumdarts}
\bibfield{author}{\bibinfo{person}{Wenjie Wu}, \bibinfo{person}{Ge Yan}, \bibinfo{person}{Xudong Lu}, \bibinfo{person}{Kaisen Pan}, {and} \bibinfo{person}{Junchi Yan}.} \bibinfo{year}{2023}\natexlab{}.
\newblock \showarticletitle{QuantumDARTS: Differentiable Quantum Architecture Search for Variational Quantum Algorithms}. In \bibinfo{booktitle}{\emph{Proceedings of the 40th International Conference on Machine Learning}} (Honolulu, Hawaii, USA) \emph{(\bibinfo{series}{ICML'23})}. \bibinfo{publisher}{JMLR.org}, Article \bibinfo{articleno}{1573}, \bibinfo{numpages}{20}~pages.
\newblock


\bibitem[Ye and Chen(2021)]%
        {ye2021quantum}
\bibfield{author}{\bibinfo{person}{Esther Ye} {and} \bibinfo{person}{Samuel Yen-Chi Chen}.} \bibinfo{year}{2021}\natexlab{}.
\newblock \showarticletitle{Quantum Architecture Search via Continual Reinforcement Learning}.
\newblock \bibinfo{journal}{\emph{arXiv preprint arXiv:2112.05779}} (\bibinfo{year}{2021}).
\newblock


\bibitem[Zhang et~al\mbox{.}(2021)]%
        {zhang2021neural}
\bibfield{author}{\bibinfo{person}{Shi-Xin Zhang}, \bibinfo{person}{Chang-Yu Hsieh}, \bibinfo{person}{Shengyu Zhang}, {and} \bibinfo{person}{Hong Yao}.} \bibinfo{year}{2021}\natexlab{}.
\newblock \showarticletitle{Neural predictor based quantum architecture search}.
\newblock \bibinfo{journal}{\emph{Machine Learning: Science and Technology}} \bibinfo{volume}{2}, \bibinfo{number}{4} (\bibinfo{year}{2021}), \bibinfo{pages}{045027}.
\newblock


\bibitem[Zhang et~al\mbox{.}(2022)]%
        {zhang2022differentiable}
\bibfield{author}{\bibinfo{person}{Shi-Xin Zhang}, \bibinfo{person}{Chang-Yu Hsieh}, \bibinfo{person}{Shengyu Zhang}, {and} \bibinfo{person}{Hong Yao}.} \bibinfo{year}{2022}\natexlab{}.
\newblock \showarticletitle{Differentiable quantum architecture search}.
\newblock \bibinfo{journal}{\emph{Quantum Science and Technology}} \bibinfo{volume}{7}, \bibinfo{number}{4} (\bibinfo{year}{2022}), \bibinfo{pages}{045023}.
\newblock


\bibitem[Zlokapa and Gheorghiu(2019)]%
        {zlokapa2020deep}
\bibfield{author}{\bibinfo{person}{Alexander Zlokapa} {and} \bibinfo{person}{Alexandru Gheorghiu}.} \bibinfo{year}{2019}\natexlab{}.
\newblock \showarticletitle{A Deep Learning Model for Noise Prediction on Near‑Term Quantum Devices}. In \bibinfo{booktitle}{\emph{Proceedings of the ACM/IEEE International Conference for High Performance Computing, Networking, Storage, and Analysis (SC ’19)}}. \bibinfo{publisher}{Association for Computing Machinery and IEEE Computer Society}, \bibinfo{address}{Denver, CO, USA}, \bibinfo{pages}{1--6}.
\newblock
\href{https://doi.org/10.1145/nnnnnnn.nnnnnnn}{doi:\nolinkurl{10.1145/nnnnnnn.nnnnnnn}}
\newblock
\shownote{K. Alexander Zlokapa Master's thesis version available as arXiv:2005.10811}.


\end{thebibliography}

\appendix
\section{Settings}
\label{settings}
Experiments settings:
\begin{table}[htbp]
     \centering
     \caption{Basic settings in different experiments.}
     \begin{threeparttable}
         \begin{tabular}[width=1.0\linewidth]
         {C{1.5cm}C{0.9cm}C{1.8cm}c}
             \hline Experiments & \#Qubits & \#Placeholders & \#Param. blocks$^*$ \\
             \hline Max-cut & 8 & 4 & 1 \\
                    JSSP & 5 & 4 & 1 \\
                    QC$^*$ & 4/\ 6& 4 &1/\ 2 \\
                    Fidelity & 3 & 6 & 1 \\
             \hline      
         \end{tabular}
         \begin{tablenotes}
             \item[*] \#Param. blocks indicate the number of parameterized blocks used in the training process.
             \item[*] QC stands for quantum chemistry
         \end{tablenotes}
     \end{threeparttable}
     \label{Tab:setting}
 \end{table}

Operation pools used for study the fidelity:
\begin{equation}
    \begin{aligned}
        \mathcal{O}_{f1} = \{&\texttt{X}:[0],\texttt{X}:[1],\texttt{X}:[2], \\
        &\texttt{T}:[0],\texttt{T}:[1],\texttt{T}:[2],, \\
        &\texttt{E}:[0,1,2] \\
        \mathcal{O}_{f2} = \{&\texttt{X}:[0],\texttt{X}:[1],\texttt{X}:[2],\\
        &\texttt{T}:[0],\texttt{T}:[1],\texttt{T}:[2],\\
        &\texttt{CNOT}:[0], \texttt{CNOT}:[1], \texttt{CNOT}:[2], \texttt{CNOT}:[0,1,2], \\
        &\texttt{CZ}:[0], \texttt{CZ}:[1], \texttt{CZ}:[2], \texttt{CZ}:[0,1,2],\\
        &,\texttt{E}:[0,1,2,3,4]\}
    \end{aligned}
\end{equation}

\section{Experiments on Max-cut Problem}
\label{Appendix Maxcut}
Baseline in Max-cut problem:
\begin{figure}[!ht]
    \centering
    \includegraphics[width=0.7\linewidth]{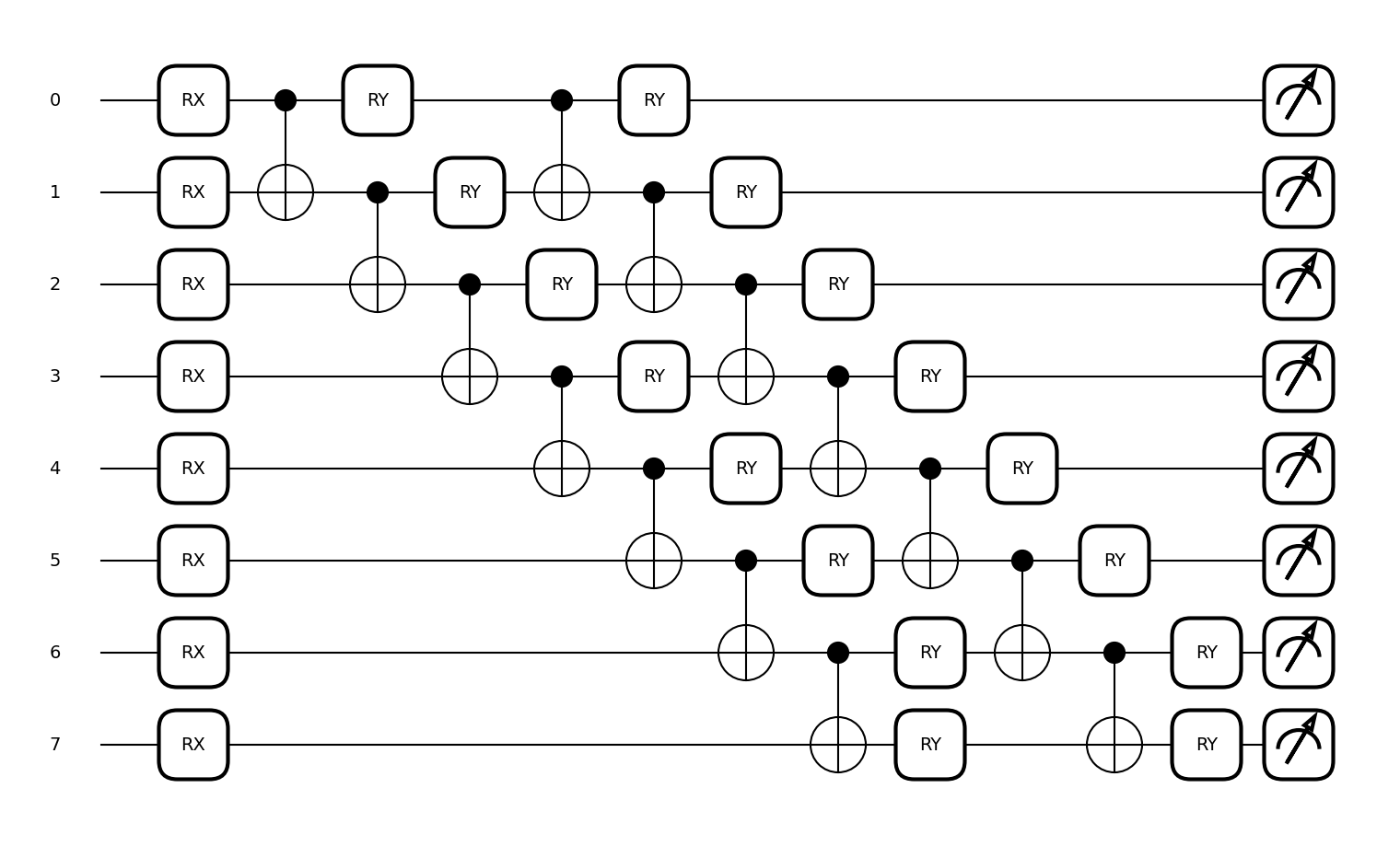}
    \caption{Baseline used for Max-cut problem.}
    \label{Fig.baseline-8}
\end{figure}
In this part, we compare the structural and performing differences between the circuits generated with SA-DQAS and DQAS and the performances in solving different Max-cut problem in the benchmark. 
\begin{table}[!ht]
     \centering
     \caption{Summary of circuit architectures designed with and without self-attention from various operation pools for solving different Max-Cut problems. M$_1$ indicates circuits searched by Method 1, M$_2$ by Method 2, and N by DQAS. \#Gate, \#Param, and \#Con refer to the number of total gates, controlled gates, and trainable parameterized gates in the parameterized blocks, respectively. R, L, and B represent Random, Ladder, and Barbell graphs, respectively.}
     \begin{threeparttable}
         \begin{tabular}[]{C{1.5cm}C{0.3cm}C{0.3cm}C{0.3cm}C{0.3cm}C{0.3cm}C{0.3cm}C{0.3cm}C{0.3cm}C{0.3cm}}
             \hline \multicolumn{1}{c}{Circuits} &
                    \multicolumn{3}{c}{\#Gate} & 
                    \multicolumn{3}{c}{\#Param} &
                    \multicolumn{3}{c}{\#Con} \\
             \hline    & M$_1$ & M$_2$ & N & M$_1$ & M$_2$ & N & M$_1$ & M$_2$ & N  \\
             \hline C$_{R-\texttt{op4}-7_4}$ & 23 & 22 & 29 & 15 & 6 & 11 & 8 & 16 & 8 \\
                    C$_{L-\texttt{op3}-3_4}$ & 23 & 27 & 22 & 15 & 19 & 14 & 8 & 8 & 8 \\
                    C$_{B-\texttt{op3}-6_4}$ & 29 & 30 & 23 & 21 & 22 & 23 & 8 & 8 & 0 \\
             \hline 
         \end{tabular}
     \end{threeparttable}
     \label{Tab:summaryM1M2N}
\end{table}

According to the Table \ref{Tab:summaryM1M2N}, the circuit C$_{R-\texttt{op4}-7_4}$ from method 2 has the fewest gates, especially parameterized ones. In the Ladder graph, circuits derived from op3-3 by both methods perform similarly, converging almost simultaneously. For the Barbell graph, the DQAS-generated circuit lacks controlled gates, crucial for quantum entanglement, making it more akin to a classical circuit. In contrast, SA-DQAS circuits, though involving higher hardware costs for entanglement, maintain better quantum properties.

These comparison results indicate that integrating self-attention with DQAS leads to better circuit structures that retain quantum features through entanglement. Such circuits demonstrate greater stability, repeatability, and insensitivity to initial parameters. Additionally, the circuits generated with SA-DQAS have fewer gates, which conserves hardware resources and reduces the errors and noise caused by gates.
\section{Experiments on JSSP}
\label{jssp}
\begin{figure}[ht]
    \centering
    \subfloat[Comparison\label{fig: best structure}]{%
    \includegraphics[width=0.37\linewidth]{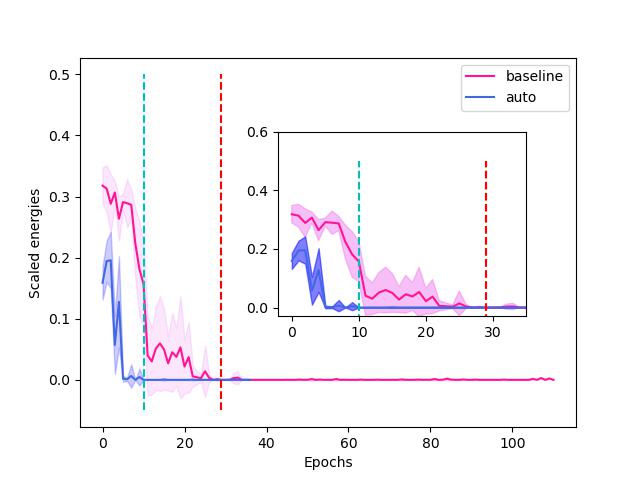}}
    \subfloat[Baseline]{
    \includegraphics[width=0.3\linewidth]{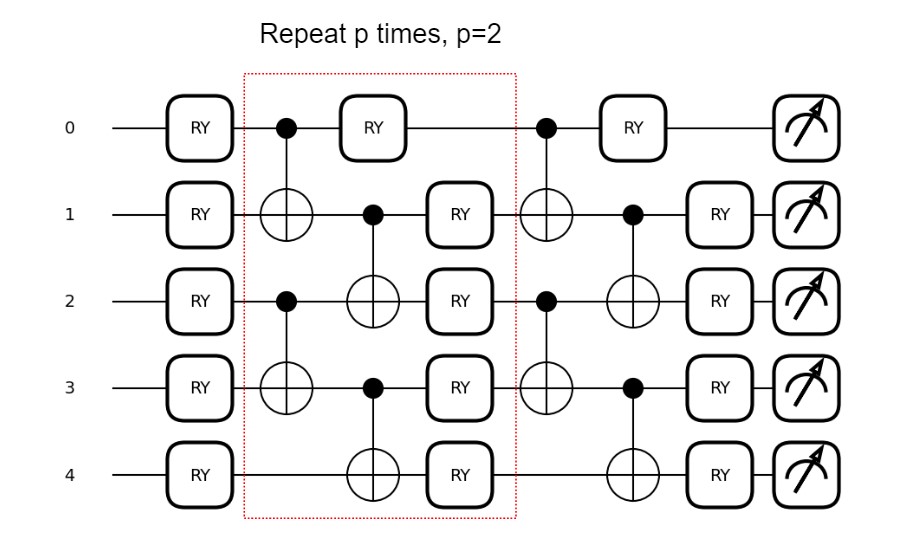}
    \label{fig: baseline}
    }
    \subfloat[Circuit$_{auto}$]{
    \includegraphics[width=0.3\linewidth]{images/bestPerformCircuit-5.jpg}
    \label{fig: best architecture}
    }
  \caption{Evaluation of newly found architectures on the simulator without noise. The results are averaged over 10 trials with the different initial parameters.}
  \label{fig: best structures}
\end{figure}
Fig. \ref{fig: best structure} presents the evaluation of newly discovered architectures on a noise-free simulator. The best-performing automatically designed architectures converge faster than the baseline to the ASP, with average results reaching the optimal schedule within 10 epochs. The shaded areas representing standard deviation are much smaller than the baseline, indicating that these circuits produce consistent outcomes across multiple trials, reflecting the stability, reliability, and reproducibility of the learned architectures. Comparing the top-performing architecture created by SA-DQAS in Fig. \ref{fig: best architecture} with the baseline circuit in Fig. \ref{fig: baseline}, the former contains no redundant parameterized blocks and ends with repeated \texttt{cnot} gates over the range [0,1,2,3], a configuration that rarely appears in manually designed circuits.

In general, more gates, especially controlled gates, increase the cost of building a quantum circuit, although controlled gates are essential for creating quantum entanglement. The circuit discovered by SA-DQAS contains fewer gates than the baseline, particularly fewer parameterized gates, reducing the building cost while retaining quantum features. Fewer parameterized gates simplifies the quantum computation process, leading to faster convergence. Additionally, since some noise and errors in quantum circuits are caused by decoherence and gate imperfections, using fewer gates can minimize these issues, resulting in more reliable and stable circuits that converge faster.

\begin{figure}[t!]
    \centering
    \subfloat[Op*-01\label{fig:rx op1}]{%
        \includegraphics[width=0.47\linewidth]{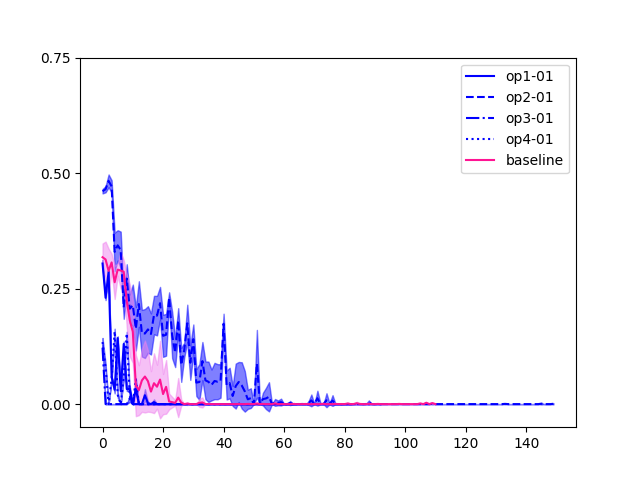}}
    \quad
    \subfloat[Op*-02\label{fig:rx op2}]{%
       \includegraphics[width=0.47\linewidth]{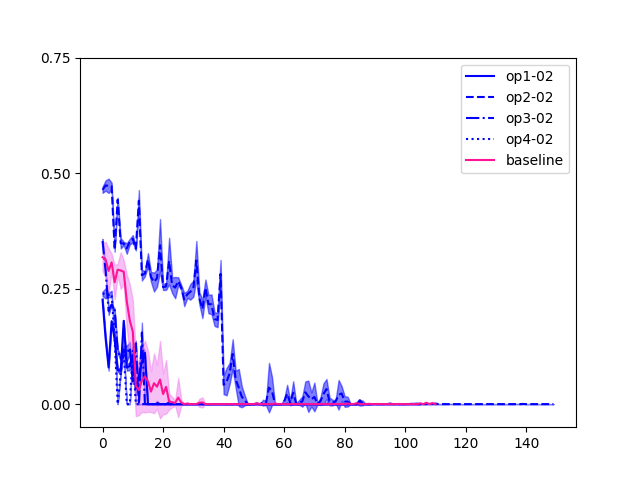}}
    \\
    \subfloat[Op*-03\label{fig:rx op3}]{%
       \includegraphics[width=0.47\linewidth]{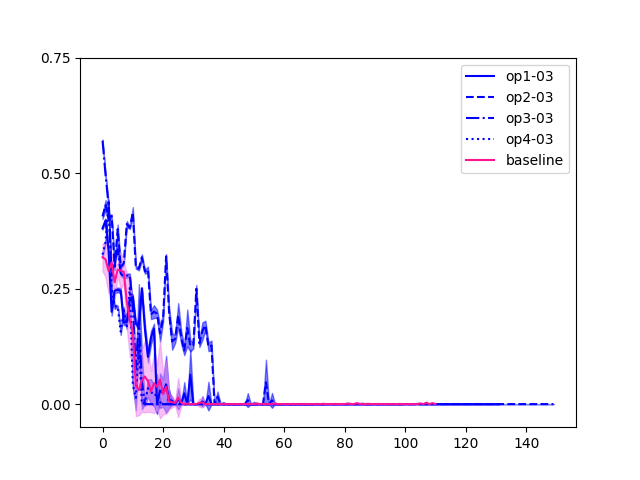}}
   \quad
   \subfloat[Op*-04\label{fig:op4}]{%
      \includegraphics[width=0.47\linewidth]{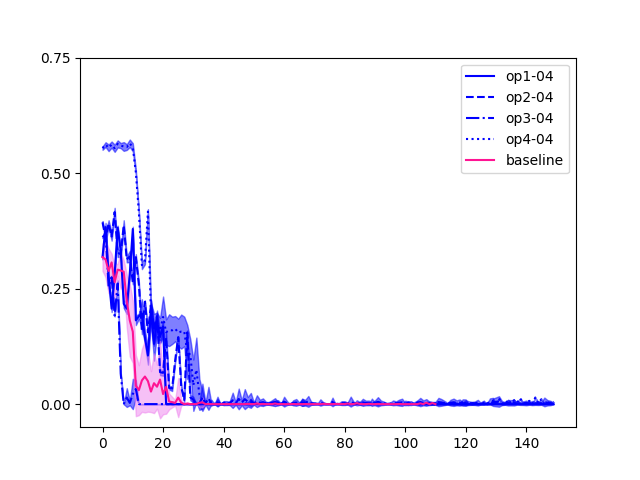}}
  \caption{Comparison of circuits generated using operation pools replacing ry with rx. We also generate operation pools by replacing ry with rx gates.}
  \label{fig:comparedqas-rx}
\end{figure}

    We also replace the all Ry gates with Rx gates in all operation pools and comparing them in JSSP problem in Fig~\ref{fig:comparedqas-rx}. Not all operation pools can find a better performing circuit architecture comparing with the baseline. But all of the circuit performance are much more stable than the baseline. The circuit deriving from op3-04 shown in subfigure~\ref{fig:op4} has the best performance comparing with other operation pools.

\section{Experiment of Fidelity Measurement}
\label{fidelity}
In~\cite{zhang2022differentiable}, only single-qubit operations were used in the operation pool. This approach helps mitigate quantum errors by converting coherent errors into more manageable Pauli errors~\cite{wallman2016noise,zhang2022differentiable,zlokapa2020deep}. We extend this by adding controlled gates to form a new operation pool, $\mathcal{O}_{f2}$ (detailed in Appendix \ref{settings}). This allows the automatically designed parameterized blocks in the circuit architectures to include entangled gates. Using both the original single-qubit pool and the new pool, we generate circuits under ideal conditions and with 20\% BitFlip noise added to each qubit at the end of the circuit. The average fidelity of the generated circuits is calculated under these conditions, and the comparisons are presented in Table \ref{Tab:summaryB1B2}.
\begin{table}[!ht]
    \centering
    \caption{A table that records the fidelity between the generated circuits from operation pools with and without controlled gates and the ideal circuit under different environments, with placeholders positioned in three locations relative to the \texttt{QFT} part. Y indicates the operation pool contains controlled gates, while N indicates only single-qubit gates are included in the operation pool.}
    \begin{threeparttable}
        \begin{tabular}[width=1.0\textwidth]{C{2.2cm}C{0.4cm}C{0.4cm}C{0.4cm}C{0.4cm}C{0.4cm}C{0.4cm}C{0.4cm}}
            \hline  & & \multicolumn{2}{c}{Ideal} & \multicolumn{2}{c}{BitFlip 2}
                    & \multicolumn{2}{c}{BitFlip 3}\\
            \hline & & Y & N & Y & N & Y & N \\
            \hline Back & B$_{0.0}$ & 0.91 & \textcolor{green}{0.94} & 0.66 & \textcolor{green}{0.68} & 0.56 & \textcolor{green}{0.65} \\
                        & B$_{0.2}$ & 0.94 & \textcolor{green}{0.97} & 0.67 & \textcolor{green}{0.69} & 0.63 & \textcolor{green}{0.65} \\
            \hline Back and Front & B$_{0.0}$ & 0.89 & \textcolor{green}{0.97} & 0.63 & \textcolor{green}{0.68} & 0.53 & \textcolor{green}{0.66} \\
                        & B$_{0.2}$ & 0.89 & \textcolor{green}{0.94} & 0.64 & \textcolor{green}{0.67} & 0.51 & \textcolor{green}{0.65} \\
            \hline Front & B$_{0.0}$ & 0.99 & 0.99 & 0.65 & 0.65 & 0.50 & \textcolor{green}{0.64} \\
                        & B$_{0.2}$ & 0.99 & 0.99 & 0.64 & \textcolor{green}{0.66} & 0.47 & \textcolor{green}{0.64} \\
            \hline
        \end{tabular}
    \end{threeparttable}
    \label{Tab:summaryB1B2}
\end{table}

When circuits (excluding the \texttt{QFT} part) contain controlled gates, the average fidelity is lower than circuits with only single-qubit gates, regardless of whether they are in ideal or noisy environments. Controlled gates link multiple qubits, so when one qubit, particularly the control qubit, is affected by an error, it impacts the state of the other qubits. This makes controlled gates more susceptible to noise, leading to lower fidelity in the final circuit. Additionally, circuits with more controlled gates can result in longer idle times for qubits, increasing the likelihood of idling errors.

To further study error mitigation in various noisy environments, we design additional noise models by introducing 20\% of different types of errors at the end of each qubit. In our new configurations, the six placeholders can be placed either all before, all after, or half before and half after the \texttt{QFT} part. We continue to use the same operation pool as in the previous experiment. Circuits are automatically generated under both ideal and noisy conditions, and we calculate the fidelity between the ideal circuit and the SA-DQAS-generated circuit in different environments.

\begin{table}[!ht]
    \centering
    \caption{A table that records the fidelity of the generated circuit under different environments and the baseline when the placeholders are all behind the \texttt{QFT} part.}
    \begin{threeparttable}
        \begin{tabular}[width=1.0\textwidth]{C{1.5cm}C{1.2cm}C{1.2cm}C{1.2cm}C{1.2cm}}
            \hline   & Ideal & P$_{0.2}$ & A$_{0.2}$ & D$_{0.2}$ \\
            \hline Ideal & 0.99 & 0.85 & 0.85 & 0.65 \\
                   P$^{*}_{0.2}$ & 0.99 & 0.85 & 0.85 & 0.81 \\
                   A$^{*}_{0.2}$ & 0.97 & 0.83 & 0.83 & 0.63 \\
                   D$^{*}_{0.2}$ & 0.99 & 0.85 & 0.85 & 0.65 \\
            \hline
        \end{tabular}
        \begin{tablenotes}
            \item[*] P indicates PhaseDamping, A is AmplitudeDamping and D is DepolarizingChannel.
                      Their subscripts indicate the probability of noise occurrence, 0.2 is 20\%.
        \end{tablenotes}
    \end{threeparttable}
    \label{Tab:PADBehind}
\end{table}

\begin{table}[!ht]
    \centering
    \caption{A table that records the fidelity of the generated circuit under different environments and the baseline when the placeholders are all before the \texttt{QFT} part.}
    \begin{threeparttable}
        \begin{tabular}[width=1.0\textwidth]{C{1.5cm}C{1.2cm}C{1.2cm}C{1.2cm}C{1.2cm}}
            \hline   & Ideal & P$_{0.2}$ & A$_{0.2}$ & D$_{0.2}$ \\
            \hline Ideal & 0.99 & 0.85 & 0.85 & 0.65 \\
                   P$_{0.2}$ & 0.99 & 0.85 & 0.85 & 0.65 \\
                   A$_{0.2}$ & 0.99 & 0.85 & 0.85 & 0.65 \\
                   D$_{0.2}$ & 0.99 & 0.85 & 0.85 & 0.65 \\
            \hline
        \end{tabular}
    \end{threeparttable}
    \label{Tab:PADBefore}
\end{table}

\begin{table}[!ht]
    \centering
    \caption{A table that records the fidelity of the generated circuit under different environments and the baseline when the placeholders are half before and half behind the \texttt{QFT} part.}
    \begin{threeparttable}
        \begin{tabular}[width=1.0\textwidth]{C{1.5cm}C{1.2cm}C{1.2cm}C{1.2cm}C{1.2cm}}
            \hline   & Ideal & P$_{0.2}$ & A$_{0.2}$ & D$_{0.2}$ \\
            \hline Ideal & 0.99 & 0.85 & 0.85 & 0.65 \\
                   P$_{0.2}$ & 0.99 & 0.85 & 0.85 & 0.65 \\
                   A$_{0.2}$ & 0.99 & 0.85 & 0.85 & 0.65 \\
                   D$_{0.2}$ & 0.99 & 0.85 & 0.85 & 0.65 \\
            \hline
        \end{tabular}
    \end{threeparttable}
    \label{Tab:PADBoth}
\end{table}

Focusing on the results in each table, we observe that the average fidelity between the ideal circuit and the generated circuits, calculated under specific environments, remains nearly the same regardless of whether the circuit was searched in an ideal or noisy environment. The fidelity is close to 1 in the ideal environment, indicating that the generated circuit closely matches the baseline. Comparing the results across the three tables, we find that, regardless of the \texttt{QFT} part's position, SA-DQAS consistently discovers circuits nearly identical to the baseline, with similar fidelity results. The lowest fidelity is over 60\%, demonstrating that SA-DQAS is capable of searching for circuit architectures in noisy environments and finding circuits that are almost identical to the baseline. Furthermore, the automatically generated circuits exhibit strong noise resistance.

\end{document}